\RequirePackage{etex}
\documentclass[seceqn,secthm]{elsart}

\usepackage{mathrsfs, epsfig}
\usepackage{dcpic, pictex, pinlabel, rotating} %setspace, amsthm
\usepackage{array, mathdots, bbm, faktor}

\usepackage{latexsym}
\usepackage{amssymb}
\usepackage{amsmath}
\usepackage{stmaryrd}
\usepackage{graphicx}
\usepackage{pst-all} %PStricks package
\usepackage{times}

\numberwithin{equation}{section}
\numberwithin{figure}{section}
\numberwithin{table}{section}

\newcommand{\bra}[1]{\langle {#1} |}
\newcommand{\ket}[1]{| {#1} \rangle}

\newcommand{\R}{{\mathbb R}}

\journal{arXiv}

\begin{document}

\begin{frontmatter}

\title{Twisted Interferometry: the topological perspective}

\author[StationQ]{Parsa Bonderson},
\author[StationQ]{Lukasz Fidkowski},
\author[StationQ,UCSBMath]{Michael Freedman},
\author[StationQ]{Kevin Walker}
\address[StationQ]{Station Q, Microsoft Research, Santa Barbara, California 93106-6105, USA}
\address[UCSBMath]{Department of Mathematics, University of California, Santa Barbara, California 93106, USA}

\begin{abstract}
Three manifold topology is used to analyze the effect of anyonic interferometers in which the probe anyons' path along an arm crosses itself, leading to a ``twisted'' or braided space-time trajectory for the probe anyons. In the case of Ising non-Abelian anyons, twisted interferometry is shown to be able to generate a topologically protected $\pi/8$-phase gate, which cannot be generated from quasiparticle braiding.
\end{abstract}

\date{29 January 2016}

\end{frontmatter}

%03.65.Ud 	  Entanglement and quantum nonlocality
%03.65.Vf     Phases: geometric; dynamic or topological
%03.65.Yz     Decoherence; open systems; quantum statistical methods
%03.67.Lx 	  Quantum computation architectures and implementations
%03.67.Pp 	  Quantum error correction and other methods for protection against decoherence
%05.30.Pr     Fractional statistics systems (anyons, etc.)
%71.10.Pm     Fermions in reduced dimensions (anyons, composite fermions,Luttinger liquid, etc.)
%73.43.-f     Quantum Hall effects

%%%%%%%%%%%%%%%%%%%%%%%%%%%%%%%%%%%%%%%%%%%%%%%%%%%%%%%%
\section{Introduction}
\label{sec:introduction}
%%%%%%%%%%%%%%%%%%%%%%%%%%%%%%%%%%%%%%%%%%%%%%%%%%%%%%%%

Anyonic interferometry~\cite{Bonderson07b,Bonderson07c} is a powerful tool for processing topological quantum information~\cite{Kitaev03,Freedman98,Preskill98,Freedman02a,Freedman02b,Freedman03b,Nayak08}. Its ability to non-demolitionally measure the collective anyonic charge of a group of (non-Abelian) anyons, without decohering their internal state, allows it to generate braid operators~\cite{Bonderson08a,Bonderson08b}, generate entangling gates~\cite{Bravyi00,Bravyi06,BondersonWIP,Levaillant2015d}, and change between different qubit encodings~\cite{BondersonWIP,Levaillant2015d}. Anyonic interferometry has been the focus of myriad experimental proposals~\cite{Chamon97,Fradkin98,DasSarma05,Stern06a,Bonderson06a,Bonderson06b,Fidkowski07c,Ardonne08a,Bishara08a,Bishara09,Akhmerov09a,Fu09a,Bonderson11b,Grosfeld11a} and efforts to physically implement them~\cite{Camino05a,Willett09a,Willett09b,McClure12,An11,Willett12a,Willett12b}. As powerful as anyonic interferometry may be, its potential capabilities have yet to be fully understood. In this paper, we propose and analyze a novel implementation of anyonic interferometry that we call ``twisted interferometry,'' which can significantly augment its potential capabilities.

One of the primary practical motivations for studying twisted interferometry is that it could be used with anyons of the Ising TQFT to generate ``magic states,'' as we will demonstrate. This is significant because, if one only has the ability to perform braiding operations and \emph{untwisted} anyonic interferometry measurements for Ising anyons, then one can only generate the Clifford group operations, which is not computationally universal and, in fact, can be efficiently simulated on a classical computer~\cite{Gottesman98}. However, if one supplements these operations with magic states, then one can also generate $\pi / 8$-phase gates, which results in a computationally universal gate set~\cite{Boykin99}.

The application of twisted interferometry to generating the $\pi / 8$-phase gate for Ising anyons is the latest link in a chain of ideas~\cite{Bravyi00-unpublished,Freedman06a,FNW05b,Bonderson10}, originating with the unpublished work of Bravyi and Kitaev, for generating a topologically-protected computational universal gate set from the Ising TQFT by utilizing topological operations. The concept and analysis of twisted interferometry is new, but closely connected to these ideas, which stem from the concept of Dehn surgery on $3$-manifolds. As we will discuss in detail, anyonic interferometry: 1) projectively measures the topological charge inside $\gamma$, and 2) decoheres the anyonic entanglement between the subsystems inside and outside the interference loop $\gamma$~\cite{Bonderson07a}. Both operations have a 3D topological interpretation in the context of Chern-Simons theory or, more generally, axiomatic (2+1)D topological quantum field theories (TQFTs). We learned from Witten~\cite{Witten89} that all low energy properties of systems governed by a TQFT can be calculated in a Euclidean signature diagrammatic formalism called unitary modular tensor categories (UMTC). This suggests~\cite{Freedman06a,FNW05b} that the choice of interference loop $\gamma$ should not be restricted to a simple space-like loop in a spatial slice $\R^2 \subset \R^2\times$time, as is the typical design for an interferometer, but rather $\gamma$ might be a general simple closed curve of space-time. Twisted interferometry explores this direction by allowing the probe anyons' path through the arms of the interferometer to be self-crossing in $\R^2$ (so $\gamma$ is \emph{immersed} in mathematical terminology). We give a general procedure for analyzing interferometers of this kind. In the restricted case of the Ising TQFT, we describe a twisted interferometer which would be capable of producing magic states.

Our strategy is: 1) to start with the UMTC calculation~\cite{Bonderson07b,Bonderson07c} which lays bare the asymptotic behavior of the simplest anyonic Mach-Zehnder interferometer (and serves as a model for Fabrey-P\'{e}rot type interferometers in the weak tunneling limit); 2) describe this behavior in an equivalent topological language; and 3) exploit the general covariance inherent in the topological description.

The concrete calculation using the machinery of UMTCs is carried out in a companion paper~\cite{Bonderson13b}, which also focuses on possible physical implementations of twisted interferometers. The analysis of the companion paper agrees with the topological argument presented here and both show how magic state production is achieved when specialized to the Ising theory.

%%%%%%%%%%%%%%%%%%%%%%%%%%%%%%%%%%%%%%%%%%%%%%%%%%%%%%%%
\section{What an Anyonic Interferometer Does in Two Different Languages}
\label{sec:what_it_does}
%%%%%%%%%%%%%%%%%%%%%%%%%%%%%%%%%%%%%%%%%%%%%%%%%%%%%%%%

We recall the bare bones of anyonic interferometry in a general anyonic context (as developed in~\cite{Bonderson07b,Bonderson07c}; see~\cite{Bonderson13b} for notational clarification and calculational details).

%%%%%%%%%%%%%%%%%%%%%%%%%%%%%%%%%%%%%%%%%%%%%%%%%%%%%%%%%%%%%%%%%%%%%%%%%%%%%
\begin{figure}[t!]
\begin{center}
  \includegraphics[scale=0.4]{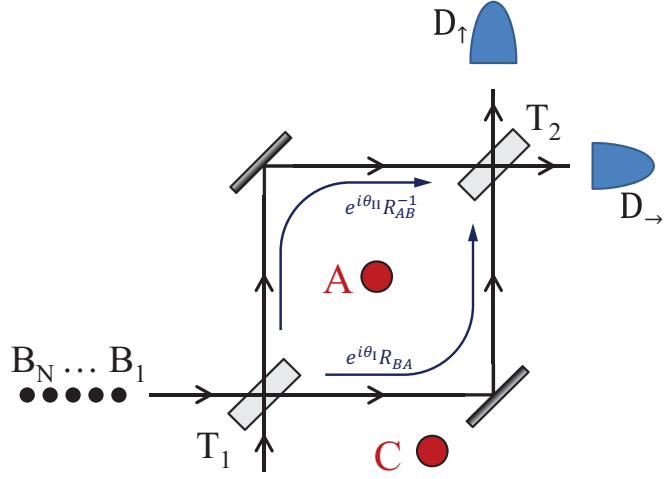}
  \caption{An idealized Mach-Zehnder interferometer for an anyonic system, where $T_{j}$ are beam splitters. The target anyons (collectively denoted $A$) in the central region share entanglement only with the anyon(s) $C$ outside this region. A beam of probe anyons $B_{1},\ldots ,B_{N}$ is sent through the interferometer and detected at one of the two possible outputs by $D_{s}$.}
  \label{fig:interferometer}
  \vspace{0.5cm}
\end{center}
\end{figure}
%%%%%%%%%%%%%%%%%%%%%%%%%%%%%%%%%%%%%%%%%%%%%%%%%%%%%%%%%%%%%%%%%%%%%%%%%%%%%

The target anyon $A$ may be a composite of several quasiparticles (anyons), so it is not necessarily in an eigenstate of charge. In the simplest case, which we treat, the probe quasiparticles $B$ are assumed to be uncorrelated, identical, and simple (not composites). In fact, to make the source standard and uncorrelated, the probes will be independently drawn from the vacuum together with an antiparticle (topological charge conjugate anyon), which is then discarded and mathematically ``traced out.'' We will simplify the discussion in this paper by also assuming the probe has definite topological charge values $B=b$, but the generalization is straightforward. Coming from the left, probe anyon $B_i$ encounters first beam splitter $T_1$, and then $T_2$.  The corresponding transition matrices are:
\begin{equation}
\label{eq:transition_matrices}
  T_{j}=\left[
    \begin{array}{cc}
      t_{j} & r_{j}^{\ast } \\
      r_{j} & -t_{j}^{\ast }
    \end{array}
  \right]
.
\end{equation}
The unitary operator representing a probe anyon passing through the
interferometer is given by%
\begin{equation}
  U = T_{2}\Sigma T_{1}
\end{equation}
\begin{equation}
  \Sigma = \left[
    \begin{array}{cc}
      0 & e^{i\theta _{\text{II}}}R_{AB}^{-1} \\
      e^{i\theta _{\text{I}}}R_{BA} & 0%
    \end{array}%
  \right] .
\end{equation}%
This can be written diagrammatically as%
\begin{equation}
\label{eq:Udiag}
\pspicture[shift=-0.4](-0.2,0)(1,1)
\rput[tl](0,0){$B_{s'}$}
\rput[tl](0,1.2){$A$}
\rput[tl](0.85,1.2){$B_{s}$}
\rput[tl](0.85,0){$A$}
 \psline[linewidth=0.9pt](0.4,0)(0.4,0.24)
 \psline[linewidth=0.9pt](0.78,0)(0.78,0.24)
 \psline[linewidth=0.9pt](0.4,0.78)(0.4,1)
 \psline[linewidth=0.9pt](0.78,0.78)(0.78,1)
\rput[bl](0.29,0.24){\psframebox{$U$}}
  \endpspicture
=e^{i\theta _{\text{I}}}\left[
\begin{array}{cc}
t_{1}r_{2}^{\ast } & r_{1}^{\ast }r_{2}^{\ast } \\
-t_{1}t_{2}^{\ast } & -r_{1}^{\ast }t_{2}^{\ast }
\end{array}
\right]_{s,s'}
\pspicture[shift=-0.4](-0.2,0)(1.25,1)
  \psset{linewidth=0.9pt,linecolor=black,arrowscale=1.5,arrowinset=0.15}
  \psline(0.92,0.1)(0.2,1)
  \psline{->}(0.92,0.1)(0.28,0.9)
  \psline(0.28,0.1)(1,1)
  \psline[border=2pt]{->}(0.28,0.1)(0.92,0.9)
  \rput[tl]{0}(-0.2,0.2){$B$}
  \rput[tr]{0}(1.25,0.2){$A$}
  \endpspicture
+e^{i\theta _{\text{II}}}\left[
\begin{array}{cc}
r_{1}t_{2} & -t_{1}^{\ast }t_{2} \\
r_{1}r_{2} & -t_{1}^{\ast }r_{2}
\end{array}
\right]_{s,s'}
\pspicture[shift=-0.4](-0.2,0)(1.25,1)
  \psset{linewidth=0.9pt,linecolor=black,arrowscale=1.5,arrowinset=0.15}
  \psline{->}(0.28,0.1)(0.92,0.9)
  \psline(0.28,0.1)(1,1)
  \psline(0.92,0.1)(0.2,1)
  \psline[border=2pt]{->}(0.92,0.1)(0.28,0.9)
  \rput[tl]{0}(-0.2,0.2){$B$}
  \rput[tr]{0}(1.25,0.2){$A$}
  \endpspicture
,
\end{equation}
where we introduce the notation of writing the directional index $s$ of the probe quasiparticle as a subscript on its anyonic (topological) charge label, e.g. $B_s$.  The anyonic state complementary to the region being probed will be denoted by $C$ (and later by two disjoint sectors $C_1$ and $C_2$).

The passage of a single probe $B$ transforms the density matrix $\rho^{AC}$ for both system and environment by
\begin{equation}
\label{eq:rhoA_single_probe}
\rho^{AC} \mapsto \rho^{AC}\left( s \right) = \frac{1}{\Pr \left( s\right) } \widetilde{\text{Tr}}_{B} \left[ \Pi _{s}  VU  \left( \rho^{B} \otimes \rho^{AC} \right)  U^{\dagger }V^{\dagger} \Pi _{s} \right],
\end{equation}
where $\widetilde{\text{Tr}}$ is the ``quantum trace,'' $V$ represents braiding, and
\begin{equation}
\text{Pr}(s) = \widetilde{\text{Tr}}[\Pi_sVU\rho U^{\dagger }V^{\dagger}]
\end{equation}
is the probability of measurement outcome $s$. The effect of this superoperator can be computed by considering the action on the $\rho^{AC}$ density matrix's basis elements, which is expressed diagrammatically by
\begin{equation}
\psscalebox{.75}{
 \pspicture[shift=-3](0.2,-3.1)(4.0,3.2)
  \small
%%%%% Boxes:
  \psframe[linewidth=0.9pt,linecolor=black,border=0](0.6,0.9)(1.7,1.4)
  \psframe[linewidth=0.9pt,linecolor=black,border=0](0.6,-0.9)(1.7,-1.4)
  \rput[bl]{0}(1.0,1.01){$U$}
  \rput[tl]{0}(0.95,-0.97){$U^{\dag}$}
%%%%% Line connections:
  \psset{linewidth=0.9pt,linecolor=black,arrowscale=1.5,arrowinset=0.15}
%  \psline(1,-0.9)(1,0.9)
  \psline(0.8,-1.4)(0.8,-2.8)
  \psline(0.8,1.4)(0.8,2.8)
  \psline(0.8,0.9)(0.8,-0.9)
  \psarc[linewidth=0.9pt,linecolor=black,border=0pt] (1.5,0.9){0.7}{-60}{0}
  \psarc[linewidth=0.9pt,linecolor=black,arrows=->,arrowscale=1.4,
    arrowinset=0.15] (1.5,0.9){0.7}{-60}{-10}
  \psarc[linewidth=0.9pt,linecolor=black,border=0pt]
(2.2,0.9){0.7}{180}{240}
  \psarc[linewidth=0.9pt,linecolor=black,arrows=<-,arrowscale=1.4,
    arrowinset=0.15] (2.2,0.9){0.7}{190}{240}
  \psarc[linewidth=0.9pt,linecolor=black,border=0pt](1.5,-0.9){0.7}{0}{60}
  \psarc[linewidth=0.9pt,linecolor=black,arrows=->,arrowscale=1.4,
    arrowinset=0.15] (1.5,-0.9){0.7}{0}{35}
  \psarc[linewidth=0.9pt,linecolor=black,border=0pt]
(2.2,-0.9){0.7}{120}{180}
  \psarc[linewidth=0.9pt,linecolor=black,arrows=<-,arrowscale=1.4,
    arrowinset=0.15] (2.2,-0.9){0.7}{145}{180}
%%%%
%%%%% Top caps
%%%%
  \psarc[linewidth=0.9pt,linecolor=black,border=0pt] (1.9,1.3){0.4}{90}{170}
  \psarc[linewidth=0.9pt,linecolor=black,border=0pt] (1.9,-1.3){0.4}{190}{270}
  \psline(1.9,1.7)(2.5,1.7)
  \psline(1.9,-1.7)(2.5,-1.7)
  \psarc[linewidth=0.9pt,linecolor=black,border=0pt](2.5,2.1){0.4}{-90}{0}
  \psarc[linewidth=0.9pt,linecolor=black,border=0pt](2.5,-2.1){0.4}{0}{90}
%%%%% Boxes:
  \psframe[linewidth=0.9pt,linecolor=black,border=0](2.6,2.1)(3.2,2.6)
  \psframe[linewidth=0.9pt,linecolor=black,border=0](2.6,-2.1)(3.2,-2.6)
  \rput[bl]{0}(2.7,2.18){$\Pi_s$}
  \rput[bl]{0}(2.7,-2.52){$\Pi_s$}
%%%%% Top caps
  \psarc[linewidth=0.9pt,linecolor=black,border=0pt](3.3,2.6){0.4}{0}{180}
  \psarc[linewidth=0.9pt,linecolor=black,border=0pt](3.3,-2.6){0.4}{-180}{0}
%%%%
  \psline(3.7,2.6)(3.7,-2.6)
%%%%
  \psline(1.85,-0.3)(1.85,0.3)
  \psline[border=2pt](2.2,-0.9)(2.2,-2.8)
  \psline[border=2pt](2.2,0.9)(2.2,2.8)
%%%%% Arrows:
  \psline{->}(0.8,-2.8)(0.8,-2.1)
  \psline{->}(0.8,-0.9)(0.8,0.12)
  \psline{->}(0.8,1.4)(0.8,2.4)
  \psline{->}(2.2,-2.8)(2.2,-2.1)
  \psline{->}(2.2,1.4)(2.2,2.4)
  \psline{->}(3.7,0.4)(3.7,-0.1)
  \psline{->}(1.9,1.69)(2.02,1.71)
  \psline{<-}(1.8,-1.69)(2.02,-1.7)
  \psline{->}(1.85,-0.3)(1.85,0.12)
%%%%% Labels:
  \rput[bl]{0}(0.7,2.9){$a$}
  \rput[tl]{0}(0.7,-2.8){$a'$}
  \rput[bl]{0}(2.1,2.9){$c$}
  \rput[tl]{0}(2.1,-2.8){$c'$}
  \rput[bl]{0}(0.3,-0.1){$b_{\shortrightarrow}$}
  \rput[bl]{0}(3.85,-0.1){$b_{s}$}
  \rput[bl]{0}(1.55,-0.2){$f$}
    \rput[bl]{0}(1.27,0.4){$a$}
  \rput[bl]{0}(1.23,-0.6){$a'$}
  \scriptsize
  \rput[bl]{0}(1.76,0.5){$\mu$}
  \rput[bl]{0}(1.71,-0.7){$\mu'$}
 \endpspicture
}
\end{equation}%

For the outcome $s=\shortrightarrow $, this may be expanded as%
\begin{eqnarray}
&& \sum\limits_{\left( e,\alpha ,\beta \right) }\left[ \left( F_{a^{\prime
}c^{\prime }}^{ac}\right) ^{-1}\right] _{\left( f,\mu ,\mu ^{\prime
}\right) \left( e,\alpha ,\beta \right) }
\nonumber \\
&& \times \left\{ \left|
t_{1}\right| ^{2}\left| r_{2}\right| ^{2}
\pspicture[shift=-1.1](0,-1.2)(1.9,1.15)
  \small
%%%%% Ellipse:
  \psellipse[linewidth=0.9pt,linecolor=black,border=0](1.19,0)(0.18,0.4)
  \psset{linewidth=0.9pt,linecolor=black,arrowscale=1.4,arrowinset=0.15}
  \psline{>-}(1.08,-0.30)(1.03,-0.10)
%%%%% Lines and arrows:
  \psset{linewidth=0.9pt,linecolor=black,arrowscale=1.5,arrowinset=0.15}
%  \psline(0.35,-0.6)(0.35,0.6)
  \psline(0.35,-0.775)(0.35,0.775)
  \psline{>-}(0.35,0.3)(0.35,0.6)
  \psline{<-}(0.35,-0.3)(0.35,-0.6)
%  \psline(1.55,-0.6)(1.55,0.6)
  \psline(1.55,-0.775)(1.55,0.775)
  \psline{>-}(1.55,0.3)(1.55,0.6)
  \psline{<-}(1.55,-0.3)(1.55,-0.6)
%  \psline(1.55,-0.1)(1.43,-0.08)
%  \psline[border=2pt](0.38,0.095)(0.71,0.04)
  \psline(1.13,-0.03)(1.55,-0.1)
  \psline[border=2pt](1.13,-0.03)(1.49,-0.09)
   \psline{-<}(0.35,0.1)(0.83,0.02)
   \psline(0.35,0.1)(0.92,0.005)
%%%%% Labels:
  \rput[bl]{0}(0.25,0.875){$a$}
  \rput[bl]{0}(1.45,0.875){${c}$}
  \rput[bl]{0}(0.25,-1.075){$a'$}
  \rput[bl]{0}(1.45,-1.075){${c'}$}
  \rput[br]{0}(0.9,0.2){$e$}
  \rput[br]{0}(0.97,-0.52){$b$}
  \scriptsize
  \rput[br]{0}(0.28,0.05){$\alpha$}
  \rput[br]{0}(1.82,-0.25){$\beta$}
  \endpspicture
\right.
+t_{1}r_{1}^{\ast }r_{2}^{\ast }t_{2}^{\ast }e^{i\left( \theta _{\text{I}}-\theta
_{\text{II}}\right) }
 \pspicture[shift=-1.1](-0.25,-1)(1.8,1.35)
  \small
%%%%% Ellipse:
  \psellipse[linewidth=0.9pt,linecolor=black,border=0](0.35,0.6)(0.4,0.18)
  \psset{linewidth=0.9pt,linecolor=black,arrowscale=1.4,arrowinset=0.15}
  \psline{->}(0.1,0.725)(0.3,0.77)
%%%%% Lines and arrows:
  \psset{linewidth=0.9pt,linecolor=black,arrowscale=1.5,arrowinset=0.15}
  \psline(0.35,-0.6)(0.35,0.69)
  \psline(0.35,0.84)(0.35,0.95)
  \psline[border=3pt]{>-}(0.35,0.3)(0.35,0.6)
  \psline{<-}(0.35,-0.3)(0.35,-0.6)
  \psline(1.35,-0.6)(1.35,0.95)
  \psline{>-}(1.35,0.3)(1.35,0.6)
  \psline{<-}(1.35,-0.3)(1.35,-0.6)
  \psline(0.35,0.1)(1.35,-0.1)
   \psline{->}(1.35,-0.1)(0.7,0.03)
 %  \psline(1.55,-0.1)(0.89,0.01)
%%%%% Labels:
  \rput[bl]{0}(0.25,1.05){$a$}
  \rput[bl]{0}(1.25,1.05){${c}$}
  \rput[bl]{0}(0.25,-0.9){$a'$}
  \rput[bl]{0}(1.25,-0.9){${c'}$}
  \rput[br]{0}(0.95,0.13){$e$}
  \rput[br]{0}(-0.05,0.7){$b$}
  \scriptsize
  \rput[br]{0}(0.28,0.05){$\alpha$}
  \rput[br]{0}(1.62,-0.25){$\beta$}
 \endpspicture
\nonumber \\
&&+t_{1}^{\ast }r_{1}t_{2}r_{2}e^{-i\left( \theta _{\text{I}}-\theta _{\text{II}}\right)
}
 \pspicture[shift=-1.13](-0.25,-1.4)(1.8,1)
  \small
%%%%% Ellipse:
  \psellipse[linewidth=0.9pt,linecolor=black,border=0](0.35,-0.6)(0.4,0.18)
  \psset{linewidth=0.9pt,linecolor=black,arrowscale=1.4,arrowinset=0.15}
  \psline{-<}(0.07,-0.715)(0.27,-0.76)
%%%%% Lines and arrows:
  \psset{linewidth=0.9pt,linecolor=black,arrowscale=1.5,arrowinset=0.15}
  \psline(0.35,-0.69)(0.35,0.6)
  \psline(0.35,-0.84)(0.35,-0.95)
  \psline{>-}(0.35,0.3)(0.35,0.6)
  \psline[border=3pt]{>-}(0.35,-0.55)(0.35,-0.3)
  \psline(1.35,0.6)(1.35,-0.95)
  \psline{>-}(1.35,0.3)(1.35,0.6)
  \psline{>-}(1.35,-0.55)(1.35,-0.3)
  \psline(0.35,0.1)(1.35,-0.1)
   \psline{->}(1.35,-0.1)(0.7,0.03)
 %  \psline(1.55,-0.1)(0.89,0.01)
%%%%% Labels:
  \rput[bl]{0}(0.25,0.7){$a$}
  \rput[bl]{0}(1.25,0.7){${c}$}
  \rput[bl]{0}(0.25,-1.25){$a'$}
  \rput[bl]{0}(1.25,-1.25){${c'}$}
  \rput[br]{0}(0.95,0.13){$e$}
  \rput[br]{0}(-0.05,-0.9){$b$}
  \scriptsize
  \rput[br]{0}(0.28,0.05){$\alpha$}
  \rput[br]{0}(1.62,-0.25){$\beta$}
 \endpspicture
\left. +\left| r_{1}\right| ^{2}\left| t_{2}\right|
^{2}
\pspicture[shift=-1.1](-0.7,-1.2)(1.75,1.15)
  \small
%%%%% Ellipse:
  \psellipse[linewidth=0.9pt,linecolor=black,border=0](-0.2,0.0)(0.18,0.4)
  \psset{linewidth=0.9pt,linecolor=black,arrowscale=1.4,arrowinset=0.15}
%  \psline{>-}(-0.29,-0.35)(-0.35,-0.15)
  \psline{->}(-0.358,-0.1)(-0.365,0.1)
%%%%% Lines and arrows:
  \psset{linewidth=0.9pt,linecolor=black,arrowscale=1.5,arrowinset=0.15}
%  \psline(0.35,-0.6)(0.35,0.6)
  \psline(0.35,-0.775)(0.35,0.775)
  \psline{>-}(0.35,0.3)(0.35,0.6)
  \psline{<-}(0.35,-0.3)(0.35,-0.6)
%  \psline(1.35,-0.6)(1.35,0.6)
  \psline(1.35,-0.775)(1.35,0.775)
  \psline{>-}(1.35,0.3)(1.35,0.6)
  \psline{<-}(1.35,-0.3)(1.35,-0.6)
  \psline(0.35,0.1)(1.35,-0.1)
   \psline{->}(1.35,-0.1)(0.7,0.03)
 %  \psline(1.55,-0.1)(0.89,0.01)
%%%%% Labels:
  \rput[bl]{0}(0.25,0.875){$a$}
  \rput[bl]{0}(1.25,0.875){${c}$}
  \rput[bl]{0}(0.25,-1.075){$a'$}
  \rput[bl]{0}(1.25,-1.075){${c'}$}
  \rput[br]{0}(0.95,0.13){$e$}
  \rput[br]{0}(-0.5,-0.3){$b$}
  \scriptsize
  \rput[br]{0}(0.28,0.05){$\alpha$}
  \rput[br]{0}(1.62,-0.25){$\beta$}
  \endpspicture
\right\}
\nonumber \\
&=& d_b \sum\limits_{\substack{ \left( e,\alpha ,\beta \right)  \\ \left(
f^{\prime },\nu ,\nu ^{\prime }\right) }}
\left[ \left( F_{a^{\prime}c^{\prime }}^{ac}\right) ^{-1}\right] _{\left( f,\mu ,\mu ^{\prime}\right) \left( e,\alpha ,\beta \right) }
p_{aa^{\prime }e,b}^{\shortrightarrow}
\left[ F_{a^{\prime }c^{\prime}}^{ac}\right] _{\left( e,\alpha ,\beta \right) \left( f^{\prime },\nu ,\nu^{\prime }\right) }
 \pspicture[shift=-1.1](0,-0.85)(1.3,1.3)
 \small
  \psset{linewidth=0.9pt,linecolor=black,arrowscale=1.5,arrowinset=0.15}
  \psline{->}(0.7,0)(0.7,0.45)
  \psline(0.7,0)(0.7,0.55)
  \psline(0.7,0.55) (0.2,1.05)
  \psline{->}(0.7,0.55)(0.3,0.95)
  \psline(0.7,0.55) (1.2,1.05)
  \psline{->}(0.7,0.55)(1.1,0.95)
  \rput[bl]{0}(0.28,0.1){$f'$}
  \rput[bl]{0}(1.1,1.15){$c$}
  \rput[bl]{0}(0.1,1.15){$a$}
  \psline(0.7,0) (0.2,-0.5)
  \psline{-<}(0.7,0)(0.35,-0.35)
  \psline(0.7,0) (1.2,-0.5)
  \psline{-<}(0.7,0)(1.05,-0.35)
  \rput[bl]{0}(1.1,-0.8){$c'$}
  \rput[bl]{0}(0.1,-0.8){$a'$}
  \scriptsize
  \rput[bl]{0}(0.82,0.45){$\nu$}
  \rput[bl]{0}(0.82,-0.0){$\nu'$}
  \endpspicture
\label{eq:one_probe_analysis}
\end{eqnarray}%
where we have defined%
\begin{eqnarray}
p_{aa^{\prime }e,b}^{\shortrightarrow } &=&\left| t_{1}\right| ^{2}\left|
r_{2}\right| ^{2}M_{eb}+t_{1}r_{1}^{\ast }r_{2}^{\ast }t_{2}^{\ast
}e^{i\left( \theta _{\text{I}}-\theta _{\text{II}}\right) }M_{ab}  \notag \\
&&+t_{1}^{\ast }r_{1}t_{2}r_{2}e^{-i\left( \theta _{\text{I}}-\theta _{\text{II}}\right)
}M_{a^{\prime }b}^{\ast }+\left| r_{1}\right| ^{2}\left| t_{2}\right| ^{2}
,
\end{eqnarray}%
where $M$ is the monodromy matrix $M_{ab} = \frac{S_{ab}S_{00}}{S_{0a}S_{0b}}$ (with $S$ the modular $S$-matrix), and $\theta_{\text{I}}, \theta_{\text{II}}$ are the non-universal phases associated with traversing the interferometer via the two different paths around the interferometry region. A similar calculation for $s=\shortuparrow$ gives
\begin{eqnarray}
p_{aa^{\prime }e,b}^{\uparrow } &=&\left| t_{1}\right| ^{2}\left|
t_{2}\right| ^{2}M_{eb}-t_{1}r_{1}^{\ast }r_{2}^{\ast }t_{2}^{\ast
}e^{i\left( \theta _{\text{I}}-\theta _{\text{II}}\right) }M_{ab}  \notag \\
&&-t_{1}^{\ast }r_{1}t_{2}r_{2}e^{-i\left( \theta _{\text{I}}-\theta _{\text{II}}\right)
}M_{a^{\prime }b}^{\ast }+\left| r_{1}\right| ^{2}\left| r_{2}\right| ^{2}.
\end{eqnarray}

Thus, we have the single probe measurement probabilities
\begin{equation}
\Pr \left( s\right) =\sum\limits_{a,c,f,\mu}\rho _{\left( a,c;f,\mu \right)
,\left( a,c;f,\mu \right) }^{AC} p_{aa0,B}^{s},
\end{equation}
and post-measurement state (for outcome $s$)
\begin{eqnarray}
\rho ^{AC}\left( s \right) &=& \sum\limits_{\substack{ a,a^{\prime},c,c^{\prime },f,\mu ,\mu ^{\prime } \\ \left( e,\alpha ,\beta \right) ,\left( f^{\prime },\nu ,\nu ^{\prime }\right) }} \frac{\rho _{\left(a,c;f,\mu \right) ,\left( a^{\prime },c^{\prime };f,\mu ^{\prime }\right)}^{AC}}{\left( d_{f} d_{f^{\prime }} \right)^{1/2}}
\left[ \left( F_{a^{\prime },c^{\prime }}^{a,c}\right)^{-1}\right]_{\left( f,\mu ,\mu ^{\prime }\right) \left( e,\alpha ,\beta \right) }
\nonumber \\
& &\times \frac{p_{aa^{\prime }e,B}^{s}}{\Pr \left( s\right) }
\left[F_{a^{\prime },c^{\prime }}^{a,c}\right] _{\left( e,\alpha ,\beta \right)\left( f^{\prime },\nu ,\nu ^{\prime }\right) }
\left| a,c;f^{\prime },\nu \right\rangle \left\langle a^{\prime},c^{\prime };f^{\prime },\nu ^{\prime }\right|.
\end{eqnarray}

The next step (which we sketch very lightly here) is to compute probabilities and the effect for a stream of $N$ identical probe anyons $B$, on $\rho^{AC}$. The results are:
\begin{equation}
\Pr \left( s_{1},\ldots ,s_{N}\right) =\sum\limits_{a,c,f,\mu}\rho _{\left( a,c;f,\mu \right)
,\left( a,c;f,\mu \right) }^{AC} p_{aa0,B}^{s_{1}}\ldots
p_{aa0,B}^{s_{N}},
\end{equation}
\begin{eqnarray}
&& \rho ^{AC}\left( s_{1},\ldots ,s_{N}\right) =
\sum\limits_{\substack{ a,a^{\prime},c,c^{\prime },f,\mu ,\mu ^{\prime } \\ \left( e,\alpha ,\beta \right),\left( f^{\prime },\nu ,\nu ^{\prime }\right) }}
\frac{\rho _{\left(a,c;f,\mu \right) ,\left( a^{\prime },c^{\prime };f,\mu ^{\prime }\right)}^{AC}}{\left( d_{f} d_{f^{\prime }} \right)^{1/2}}
\left[ \left( F_{a^{\prime },c^{\prime }}^{a,c}\right)^{-1}\right]_{\left( f,\mu ,\mu ^{\prime }\right) \left( e,\alpha ,\beta \right) }
\notag \\
&&\times \frac{p_{aa^{\prime }e,B}^{s_{1}}\ldots p_{aa^{\prime }e,B}^{s_{N}}}{\Pr \left( s_{1},\ldots ,s_{N}\right) }
\left[F_{a^{\prime },c^{\prime }}^{a,c}\right] _{\left( e,\alpha ,\beta \right) \left( f^{\prime },\nu ,\nu ^{\prime }\right) }\left| a,c;f^{\prime },\nu \right\rangle \left\langle a^{\prime},c^{\prime };f^{\prime },\nu ^{\prime }\right|
.
\label{eq:rhoA_N}
\end{eqnarray}%
It is clear that the specific order of the measurement outcomes is not important, but only the total number of outcomes of each type matters, and that keeping track of only the total numbers leads to a binomial distribution.

For generic choices of interferometric parameters: $t_j, r_j, \theta_{\text{I}}$, and $\theta_{\text{II}}$, these binomial distributions will concentrate exponentially fast at distinct transmission probabilities associated with the equivalence classes of charge types $a$ where $a\equiv a^{\prime}$ if and only if $M_{a,b} = M_{a^{\prime},b}$. In the simplest cases, there is a natural choice for the probe $B$ where every $a$ is distinguished (e.g. for Ising and Fibonacci anyons one selects $b = \sigma$ and $b = \tau$, respectively), and hence the ``equivalence classes'' are singletons. In general, the probability of observing $n$ (out of $N$) probes in the $\rightarrow$ detector is:
\begin{eqnarray}
 \Pr^{\kappa}\nolimits_{N}\left( n\right) &=&\sum\limits_{\kappa}\Pr\nolimits_{A}\left( \kappa \right) \frac{N!}{n!(N-n)!}p_{\kappa}^n(1-p_{\kappa})^{N-n},\\
 \Pr\nolimits_{A}\left( \kappa \right) &=& \sum\limits_{a\in \mathcal{C}_{\kappa },c,f,\mu }\rho _{\left( a,c;f,\mu \right) ,\left( a,c;f,\mu \right) }^{AC},
\end{eqnarray}
where $\kappa$ indexes the equivalence classes $\mathcal{C}_{\kappa}$ w.r.t. probe $b$. The fraction $r = n/N$ of probes measured in the $s = \rightarrow$ detector goes to $r=p_{\kappa}$ with probability $\Pr_A(\kappa)$, and the target anyon density matrix will generically collapse onto the corresponding ``fixed states.''

The asymptotic operation $N\rightarrow \infty$ of a generically tuned anyonic interferometer converges to a fixed state of charge sector $\kappa$ with probability $\Pr_A(\kappa)$ and: 1) projects the anyonic state onto the subspace where the $A$ anyons have collective anyonic charge in $\mathcal{C}_{\kappa}$, and 2) decoheres all anyonic entanglement between subsystem $A$ and $C$ that the probes can detect. The sector $\kappa$ may be a single charge or a collection of charges with identical monodromy elements with the probes, i.e. $M_{a,B} = M_{a^{\prime},B}$ for $a,a^{\prime}\in\mathcal{C}_{\kappa}$. The anyonic entanglement between $A$ and $C$ is described in the form of anyonic charge lines connecting these subsystems, i.e. the charge lines labeled by charge $e$ in the preceding analysis, where the contribution of a diagram to the density matrix will be removed if $M_{e,B}\neq 1$. Convergence to such a fixed state is based on Gaussian statistics, therefore exponentially precise as a function of the number $N$ of probe particles.

In the simplest case, $M_{a,b} = M_{a^{\prime},b} \Rightarrow a = a^{\prime}$ and the indistinguishable equivalence classes $C_{\kappa_a}=\{a\}$ are singletons, i.e. all topological charges are distinguished. The corresponding fixed state density matrix is:
\begin{equation}
\rho _{\kappa_{a} }^{AC} = \sum\limits_{c} \frac{\Pr\nolimits _{A} \left(c | a \right)}{ d_{a}d_{c} } \,\, \mathbb{I}_{ac}
= \sum\limits_{c,f^{\prime},\nu} \frac{\Pr\nolimits _{A} \left(c | a \right)}{ d_{a}d_{c} } \,\, \left| a,c;f^{\prime },\nu \right\rangle \left\langle a,c;f^{\prime },\nu \right|,
\end{equation}
where
\begin{equation}
\Pr\nolimits _{A} \left(c | a \right) = \frac{ \sum\limits_{f,\mu}\rho _{\left(
a,c;f,\mu \right) \left( a,c;f,\mu \right)}^{A} }{\sum\limits_{c,f,\mu}\rho _{\left(
a,c;f,\mu \right) \left( a,c;f,\mu \right)}^{A}}.
\end{equation}
(The formulae for the general case can be found in~\cite{Bonderson07b,Bonderson07c}.) From this point on, we focus only on these cases where the probe distinguishes all topological charges.

%%%%%%%%%%%%%%%%%%%%%%%%%%%%%%%%%%%%%%%%%%%%%%%%%%%%%%%%%%%%%%%%%%%%%%%%%%%%%
\begin{figure}[t!]
\begin{center}
  \includegraphics[scale=0.4]{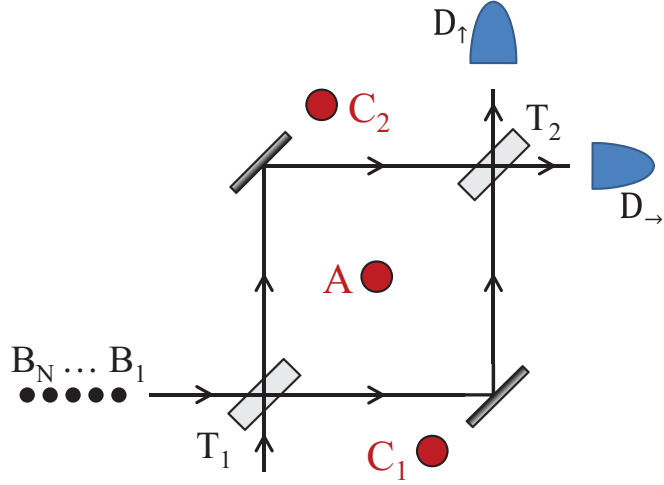}
  \caption{An idealized Mach-Zehnder interferometer where the anyons $C$ entangled with the target anyons $A$ are separated into two regions $C_1$ and $C_2$.}
  \label{fig:MZint2}
  \vspace{0.5cm}
\end{center}
\end{figure}
%%%%%%%%%%%%%%%%%%%%%%%%%%%%%%%%%%%%%%%%%%%%%%%%%%%%%%%%%%%%%%%%%%%%%%%%%%%%%

This is a convenient place to note a modest generalization, where the complementary charge $C$ is divided into two regions separated by the interferometer, which we similarly denote as $C_1$ and $C_2$, respectively. In some experimental setups --- e.g. a Fabrey-P\'{e}rot interferometer on a quantum Hall bar --- each arm of the interferometer individually will separate the region with charge $A$ from a complementary region with respective charges $C_1$ and $C_2$, which could both be nontrivial. This situation is depicted for the idealized Mach-Zehnder interferometer in Fig.~\ref{fig:MZint2}. In this circumstance, all charge lines from $A$ to $C_1$ and from $A$ to $C_2$ are (separately) decohered if they can be detected by the probes $B$.

%%%%%%%%%%%%%%%%%%%%%%%%%%%%%%%%%%%%%%%%%%%%%%%%%%%%%%%%%%%%%%%%%%%%%%%%%%%%%%%%%%%%%%%%%%%%%%%%
\begin{figure}[t!]
  \labellist
  \pinlabel $a$ at 210 650
  \pinlabel $c$ at 465 650
  \pinlabel $\alpha_2$ at 330 570
  \pinlabel $\alpha_1$ at 350 440
  \pinlabel $\alpha_4$ at 0 440
  \pinlabel $\alpha_3$ at 330 50
  \pinlabel $e$ at 420 340
  \pinlabel $a^{\prime}$ at 210 -40
  \pinlabel $c^{\prime}$ at 465 -40
  \pinlabel $\text{(a)}$ at 280 -130
  \pinlabel $c_1$ at 990 650
  \pinlabel $a$ at 1340 650
  \pinlabel $c_2$ at 1700 650
  \pinlabel $\alpha_4$ at 1090 440
  \pinlabel $\alpha_3$ at 1190 490
  \pinlabel $\alpha_2$ at 1480 490
  \pinlabel $\alpha_1$ at 1590 440
  \pinlabel $e_2$ at 1180 280
  \pinlabel $e_1$ at 1500 280
  \pinlabel $c_1^{\prime}$ at 990 -40
  \pinlabel $a^{\prime}$ at 1340 -40
  \pinlabel $c_2^{\prime}$ at 1700 -40
  \pinlabel $\text{(b)}$ at 1340 -130
  \endlabellist
\begin{center}
  \includegraphics[height=1.5in]{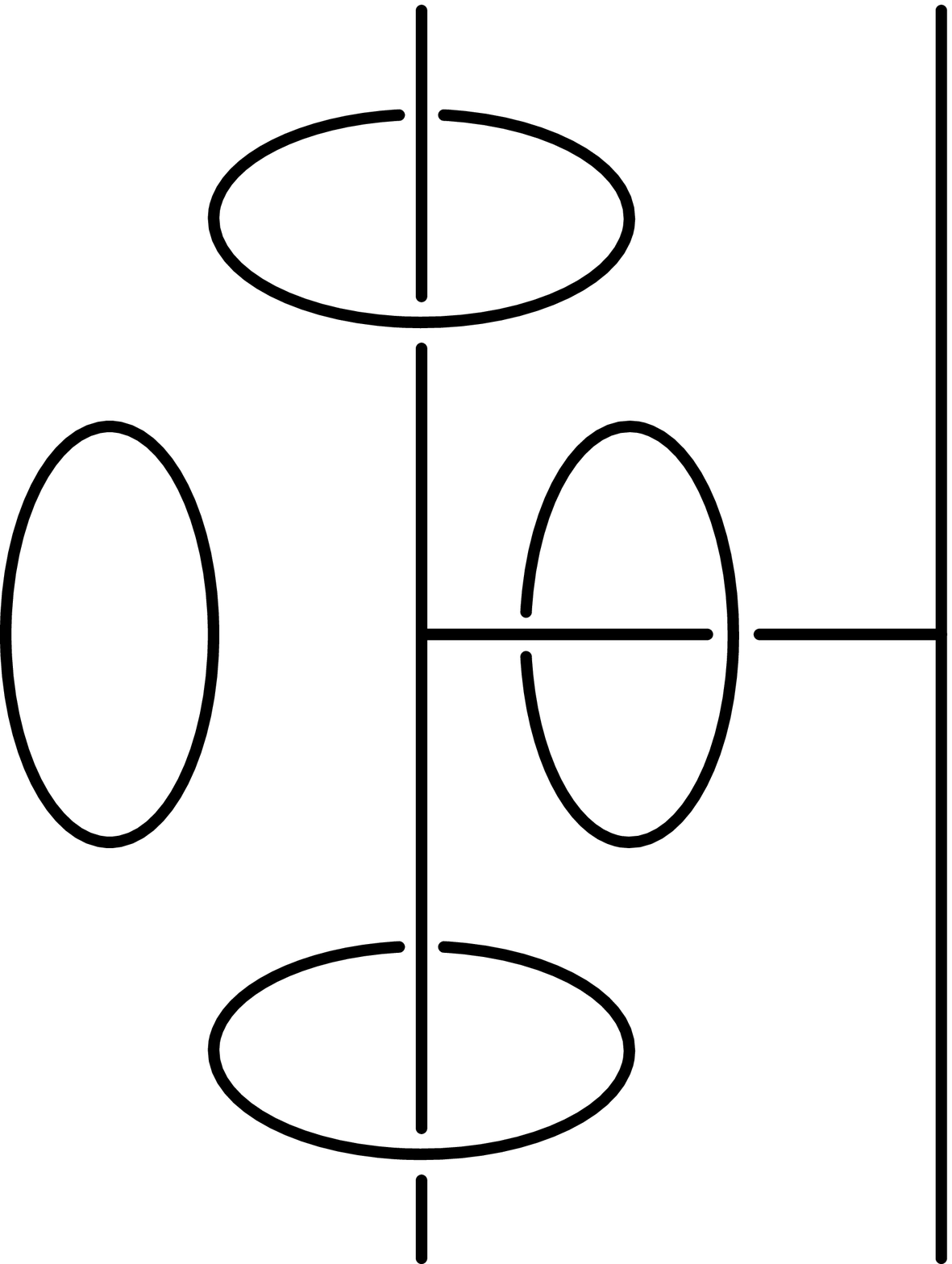}\hspace{3cm}
  \includegraphics[height=1.5in]{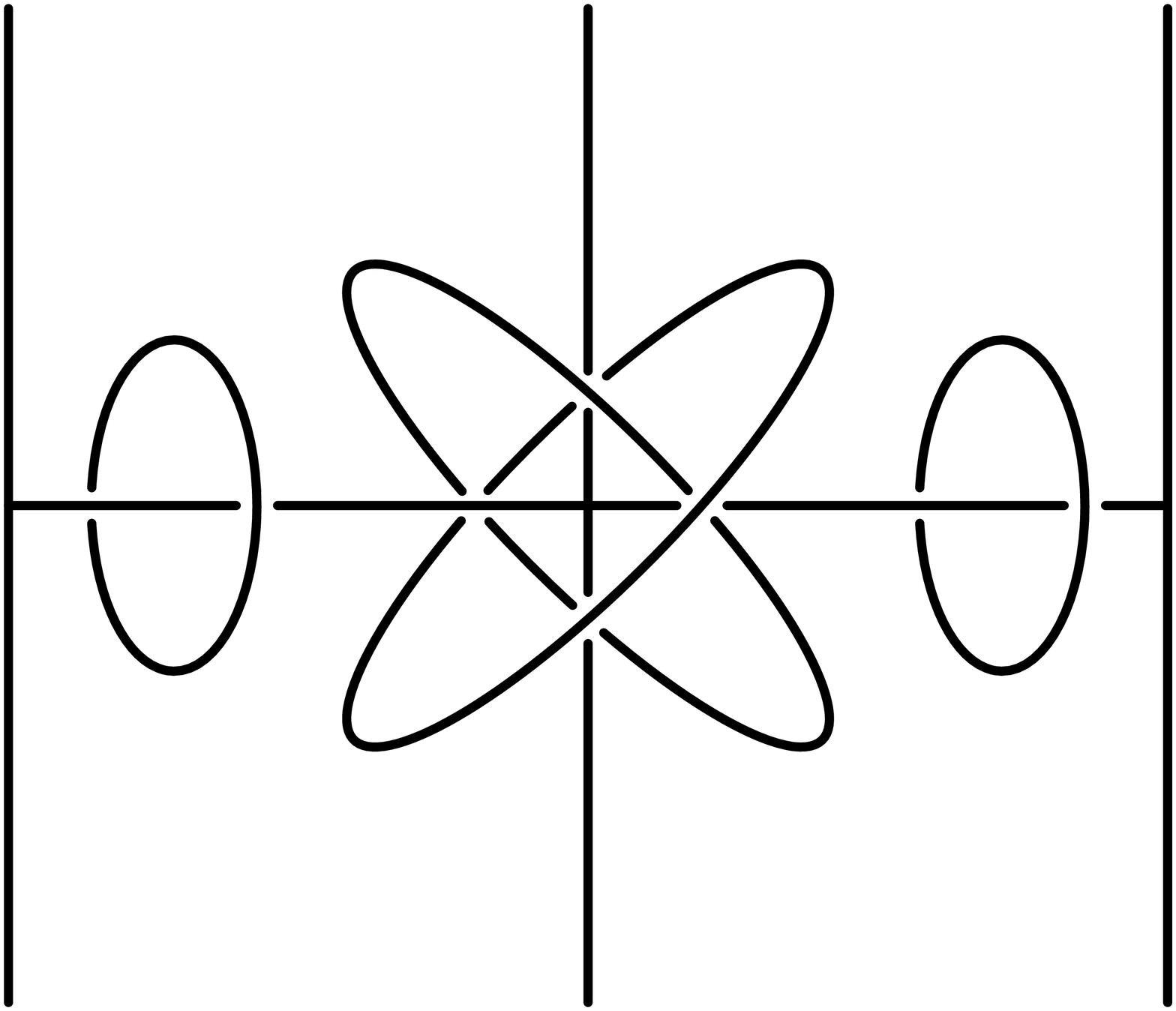}\vspace{0.75cm}
  \caption{(a) For a single region of complementary anyons $C$, we show the four positions for the probe loops corresponding to the four terms of Eq.~(\ref{eq:one_probe_analysis}). (b) For two regions of complementary charge $C_1$ and $C_2$, the four positions of probe loops are shown on the more complicated the target system (with complementary anyons) density matrix components. (The $4$-valent vertex is understood to be resolved into appropriate trivalent vertices.) $\alpha_j$ denotes the weight with which the corresponding probe loop configuration enters the measurement superoperator.}
  \label{fig:bipartite}
  \vspace{0.5cm}
\end{center}
\end{figure}

%%%%%%%%%%%%%%%%%%%%%%%%%%%%%%%%%%%%%%%%%%%%%%%%%%%%%%%%%%%%%%%%%%%%%%%%%%%%%%%%%%%%%%%%%%%%%%%%%%%%%%%%%%%%%%%%%%%%%%%%%%%

In Fig.~\ref{fig:bipartite}, we compare the diagrammatic terms that arise for a single $C$ region formulation to when there are two regions $C_1$ and $C_2$.
For probe $b$ and measurement outcome $s=\shortrightarrow$, the four probe loop configurations enter the measurement superoperator with weights
\begin{eqnarray}
\alpha_1^{\shortrightarrow} &=& |t_1|^2|r_2|^2, \\
\alpha_2^{\shortrightarrow} &=& t_1r_1^{\ast}r_2^{\ast}t_2^{\ast}e^{i(\theta_I-\theta_{II})}, \\
\alpha_3^{\shortrightarrow} &=&  t_1^{\ast}r_1t_2r_2e^{-i(\theta_I-\theta_{II})},\\
\alpha_4^{\shortrightarrow} &=& |r_1|^2|t_2|^2,
\end{eqnarray}
as in Eq.~(\ref{eq:one_probe_analysis}). For $s=\shortuparrow$, these are
\begin{eqnarray}
\alpha_1^{\shortuparrow} &=& |t_1|^2|t_2|^2, \\
\alpha_2^{\shortuparrow} &=& - t_1r_1^{\ast}r_2^{\ast}t_2^{\ast}e^{i(\theta_I-\theta_{II})}, \\
\alpha_3^{\shortuparrow} &=&  - t_1^{\ast}r_1t_2r_2e^{-i(\theta_I-\theta_{II})},\\
\alpha_4^{\shortuparrow} &=& |r_1|^2|r_2|^2.
\end{eqnarray}
Given $N$ uncorrelated identical probe anyons, there are $4^N$ configurations of probe loops, each probe choosing from the four positions, with the single probe weights (depending on a given probe's measurement outcome) being multiplied together for the overall superoperator. For the two probe loop positions which cross in Fig.~\ref{fig:bipartite}(b), repeated copies will nest according to the pattern of later probe loops having larger radius. We will see shortly that the detail of the nesting patterns are irrelevant in the large $N$ limit.

According to the calculation just summarized, the net effect of running the interferometer on the target system with density matrix $\rho^{AC}$, up to corrections that decay exponentially in $N$, is that the superposition of these $4^N$ configurations results in a measurement of the collective charge of anyons $A$ onto charge value $a$, with probability
\begin{equation}
{\Pr}_{AC}(a) = \widetilde{\text{Tr}}\left[ \rho^{AC} \Pi_{a}^{A} \right]
,
\end{equation}
and post-measurement density matrix
\begin{equation}
{\rho}_{a}^{AC} = \frac{1}{{\Pr}_{AC}(a)}
\pspicture[shift=-2.9](-2.8,-3)(2.8,2.5)
  \small
%%%%% Box:
  \psline(-2.5,-0.5)(-2.5,0.5)
  \psline(-2.5,-0.5)(2.5,-0.5)
  \psline(-2.5,0.5)(2.5,0.5)
  \psline(2.5,-0.5)(2.5,0.5)
%%%%% Line connections:
  \psset{linewidth=0.9pt,linecolor=black,arrowscale=1.5,arrowinset=0.15}
  \psline(2.0,0.5)(2.0,2)
  \psline(0.0,0.5)(0.0,2)
  \psline(2.0,-0.5)(2.0,-2)
  \psline(0.0,-0.5)(0.0,-2)
  \psline(-2.0,0.5)(-2.0,2)
  \psline(-2.0,-0.5)(-2.0,-2)
%%%%% Arrows:
  \psline{->}(2.0,1.5)(2.0,1.75)
  \psline{->}(0.0,1.5)(0.0,1.75)
  \psline{-<}(2.0,-1.5)(2.0,-1.75)
  \psline{-<}(0.0,-1.5)(0.0,-1.75)
  \psline{->}(-2.0,1.5)(-2.0,1.75)
  \psline{-<}(-2.0,-1.5)(-2.0,-1.75)
%%%%% Leg Labels:
  \rput[bl](1.9,2.05){$C_1$}
  \rput[bl](-0.1,2.1){$A$}
  \rput[bl](1.9,-2.45){$C_1^{\prime}$}
  \rput[bl](-0.1,-2.35){$A^{\prime}$}
  \rput[bl](-2.3,2.05){$C_2$}
  \rput[bl](-2.3,-2.45){$C_2^{\prime}$}
%%%% Probe Loop:
  \psbezier[linewidth=0.9pt,linecolor=black,border=0.05](-0.5,0.0)(-1.25,1.5)(-1.0,2.0)(-0.25,0.5)
  \psbezier[linewidth=0.9pt,linecolor=black,border=0.05](-0.5,0.0)(0.25,-1.5)(1.25,-2.5)(0.25,-0.5)
   \psline{<-}(-0.5,0.0)(-0.55,0.1)
  \psline[linewidth=0.9pt,linecolor=black,border=0.1](-0.4,0.5)(-0.2,0.5)
  \psline[linewidth=0.9pt,linecolor=black,border=0.1](0.4,-0.5)(0.2,-0.5)
  \psbezier[linewidth=0.9pt,linecolor=black,border=0.05](0.5,0.0)(1.5,2.0)(1.0,2.0)(0.1,0.5)
  \psbezier[linewidth=0.9pt,linecolor=black,border=0.05](0.5,0.0)(-0.5,-2.0)(-1.3,-2.0)(-0.4,-0.5)
  \psline[linewidth=0.9pt,linecolor=black,border=0.1](0.4,0.5)(0.05,0.5)
  \psline[linewidth=0.9pt,linecolor=black,border=0.1](-0.5,-0.5)(-0.35,-0.5)
   \psline{->}(0.5,0.0)(0.55,0.1)
  \psellipse[linewidth=0.9pt,linecolor=black,border=0.05](1.5,0.0)(0.3,1.5)
  \psline{->}(1.215,-0.13)(1.215,-0.1)
  \psframe[linewidth=0.9pt,linecolor=white,border=0.1,fillcolor=white,fillstyle=solid](1.7,-0.5)(1.8,0.5)
  \psline(2.0,0.5)(1.5,0.5)
  \psline(2.0,-0.5)(1.5,-0.5)
  \psellipse[linewidth=0.9pt,linecolor=black,border=0.05](-1.5,0.0)(-0.3,1.5)
  \psline{->}(-1.19,-0.13)(-1.19,-0.1)
  \psframe[linewidth=0.9pt,linecolor=white,border=0.1,fillcolor=white,fillstyle=solid](-1.7,-0.5)(-1.8,0.5)
  \psline(-2.0,0.5)(-1.5,0.5)
  \psline(-2.0,-0.5)(-1.5,-0.5)
%%%%% Labels:
  \rput[bl]{0}(-0.3,-0.15){$\rho^{AC}$}
  \scriptsize
  \rput[bl](1.35,-0.3){$\omega_{0}$}
  \rput[bl](0.6,-0.1){$\omega_{a}$}
  \rput[bl](-1.65,-0.3){$\omega_{0}$}
  \rput[bl](-0.9,-0.1){$\omega_{a}$}
 \endpspicture
\label{eq:target_projected_omega}
.
\end{equation}
(All topological charge lines drawn here have zero framing, i.e. there are no twists in the frame.)
The $\omega_a$-loops
\begin{equation}
\pspicture[shift=-0.55](-0.25,-0.1)(0.9,1.3)
\small
  \psset{linewidth=0.9pt,linecolor=black,arrowscale=1.5,arrowinset=0.15}
  \psellipse[linewidth=0.9pt,linecolor=black,border=0](0.4,0.5)(0.4,0.18)
  \psset{linewidth=0.9pt,linecolor=black,arrowscale=1.4,arrowinset=0.15}
  \psline{->}(0.2,0.37)(0.3,0.34)
  \rput[bl]{0}(0.0,0.0){$\omega_a$}
  \endpspicture
=
\pspicture[shift=-0.55](-0.25,-0.1)(0.9,1.3)
\small
  \psset{linewidth=0.9pt,linecolor=black,arrowscale=1.5,arrowinset=0.15}
  \psellipse[linewidth=0.9pt,linecolor=black,border=0](0.4,0.5)(0.4,0.18)
  \psset{linewidth=0.9pt,linecolor=black,arrowscale=1.4,arrowinset=0.15}
  \psline{-<}(0.2,0.37)(0.3,0.34)
  \rput[bl]{0}(0.0,0.0){$\omega_{\bar{a}}$}
  \endpspicture
= \sum_{x} S_{0a} S^{\ast}_{ax}
\pspicture[shift=-0.55](-0.25,-0.1)(0.9,1.3)
\small
  \psset{linewidth=0.9pt,linecolor=black,arrowscale=1.5,arrowinset=0.15}
  \psellipse[linewidth=0.9pt,linecolor=black,border=0](0.4,0.5)(0.4,0.18)
  \psset{linewidth=0.9pt,linecolor=black,arrowscale=1.4,arrowinset=0.15}
  \psline{->}(0.2,0.37)(0.3,0.34)
  \rput[bl]{0}(0.0,0.05){$x$}
  \endpspicture
\label{eq:omega_a}
\end{equation}
have the effect of projecting all charge lines passing through the loop onto collective charge $a$. Thus, the $\omega_0$-loops effectively cut charge lines. This allows the $\omega_a$-loops to be moved to encircle only the $A$ and $A'$ lines, i.e. one can perform a handle slide of the loop around the $\omega_0$-loops (see Section~\ref{sec:surgery_handle}). Thus, the $\omega_a$-loops effect projection of anyons $A$ into collective charge sector $a$. When there is only one region of complementary anyons $C$, e.g. if there are no $C_2$ anyons, then the action of the $\omega_0$-loop between $A$ and $C_2$ is trivial. Notice that the $\omega$-loops here occur in precisely the same positions as the four possible probe loop configurations.

\begin{figure}[t!]
  \labellist
  % labels for (a)
  \pinlabel $H$ at 350 650
  \pinlabel $\omega_a$ at 50 500
  \pinlabel $\omega_a$ at -50 450
  \pinlabel $\omega_0$ at 720 320
  \pinlabel $\gamma$ at 680 430
  \pinlabel $\bar{\gamma}$ at 620 210
  \pinlabel $\text{(a)}$ at 350 -110
  % labels for (b)
  \pinlabel $H^{\prime}$ at 1000 700
  \pinlabel $\omega_a$ at 1400 700
  \pinlabel $\omega_0$ at 900 330
  \pinlabel $\omega_0$ at 1830 330
  \pinlabel $\omega_a$ at 1370 -20
  \pinlabel $\beta$ at 930 480
  \pinlabel $\bar{\beta}$ at 930 180
  \pinlabel $\gamma$ at 1780 480
  \pinlabel $\bar{\gamma}$ at 1780 180
  \pinlabel $\text{(b)}$ at 1370 -110
  \endlabellist
\begin{center}
  \includegraphics[width=0.9\textwidth]{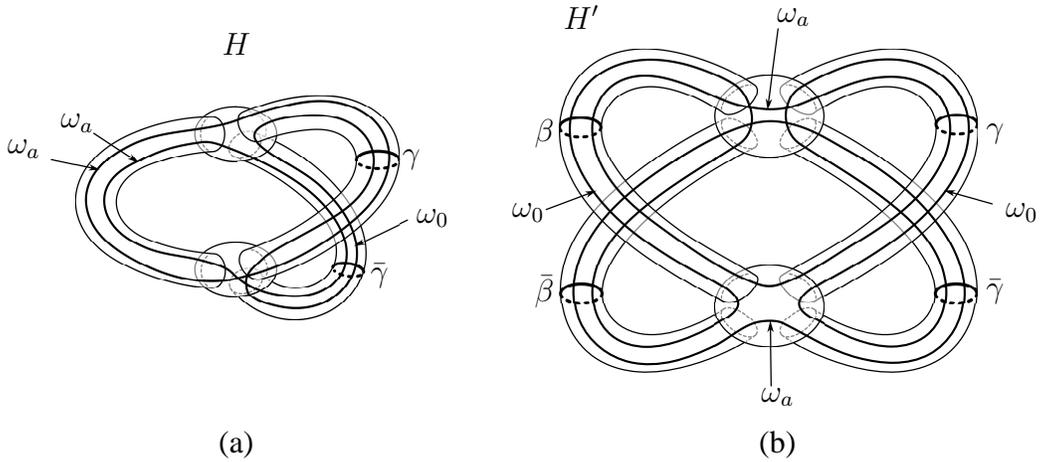}
  \vspace{0.75cm}
  \caption{(a) Genus 2 handle body $H$ and (b) genus 3 handle body $H'$, within which the effect of interferometry is localized in Eq.~(\ref{eq:target_projected_omega}). Some curves in $\partial H$ and $\partial H^{\prime}$ are labeled for later reference.}
  \label{fig:handle_bodies}
  \vspace{0.5cm}
\end{center}
\end{figure}

Having depicted the effect of interferometry in terms of $\omega$-loops, we make a geometric observation for later use: the effects of interferometry are localized to a certain quasi-1D region of space-time surrounding the $\omega$-loops called a ``handle body.'' These are indicated in Fig.~\ref{fig:handle_bodies} as the regions $H$ and $H^{\prime}$ for the single region $C$ and two region $C_1$ and $C_2$ configuration of complementary anyons. The handle-bodies $H$ and $H^{\prime}$ model the complementary regions surrounding the $\rho^{AC}$ density matrix operator. This enables us to make calculations for twisted interferometry simply by computing operators within transformed coordinates.

%%%%%%%%%%%%%%%%%%%%%%%%%%%%%%%%%%%%%%%%%%%%%%%%%%%%%%%%%%%%%%%%%%%%%%%%%%%%%
%%%%%%%%%%%%%%%%%%%%%%%%%%%%%%%%%%%%%%%%%%%%%%%%%%%%%%%%%%%%%%%%%%%%%%%%%%%%%

%%%%%%%%%%%%%%%%%%%%%%%%%%%%%%%%%%%%%%%%%%%%%%%%%%%%%%%%
\section{Topological Explanations}
\label{sec:topological_explanations}
%%%%%%%%%%%%%%%%%%%%%%%%%%%%%%%%%%%%%%%%%%%%%%%%%%%%%%%%

The goal of this section is to explain the topological nature of interferometry. In Section~\ref{sec:surgery_handle}, we first review some pure topology background on $3$-manifold surgery and the handle slide property. In Section~\ref{sec:surgical_description}, we apply this machinery to interferometry, with the basic idea being that in the limit of large $N$, the exact partition function, given by $4^N$ terms with probe anyon Wilson loops, can effectively be described by a small number of Dehn surgeries.  Although this abstract topological approach may at first seem like overkill, it proves its utility when we try to generalize to the case of twisted interferometry, which is introduced in Section~\ref{sec:twisted_interferometers}. Indeed, as shown in Section~\ref{sec:twist_consequence}, twisting has a natural description in the effective topological language: to compute the partition function in the twisted case, all we have to do is modify the gluing of a certain handle body by some twists.  Section~\ref{sec:top_understanding}, although not necessary in the logical flow of the paper, develops a stand-alone, purely topological perspective on interferometry.  Finally, in Section~\ref{sec:Ising_theory}, we apply all this machinery to the case of the Ising UMTC, and describe the simplifications that arise.

\subsection{Surgery and the Handle Slide Property}
\label{sec:surgery_handle}

``Handles'' are a combinatorial tool for assembling smooth $d$-manifolds with boundary out of little pieces, which are individually copies of $d$-balls. Our main focus is $d=4$, since we will manipulate within a $(2+1)$D TQFT using a representation where the 3D space-time is the boundary of a 4D bulk. Note, however, that the handle bodies drawn in Fig.~\ref{fig:handle_bodies} are 3D, being subsets of the space-time itself.

Let $B^d$ denote the unit ball in $\R^d$. There are $d+1$ types of $d$-dimensional handles, called $k$-handles (or handles of index $k$), where $0\leq k\leq d$. A $k$-handle is a pair $(B^k\times B^{d-k},\partial B^k\times B^{d-k})$. Note that the total space $B^k\times B^{d-k}$ is always diffeomorphic to a $d$-ball $B^d$, so what is significant is the portion of the boundary specified in the second slot. This portion is called the ``attaching region'' and consists of larger portions of $\partial B^d$ as the index $k$ increases. For example, the five $k$-handles for dimension $d=4$ are given by:
\begin{equation}
  \begin{tabular}[htpb]{c|c}
    $k$ & attaching region\\\hline
    0 & $\varnothing$ = nothing\\
    1 & $\{-1,1\}\times B^3$ = two balls\\
    2 & $S^1\times B^2$ = solid torus\\
    3 & $S^2\times I$ = spherical shell\\
    4 & $S^3$ = entire 3-sphere
  \end{tabular}
\label{tab:attaching_region}
\end{equation}
We see from this table:

0-handles are attached to nothing; they are the beginning of the construction of a 4-manifold $M^4$, corresponding to local minima of the Morse function, $x_1^2+x_2^2+x_3^2+x_4^2=0$.

1-handles attach to 0-handles, and correspond to an index = 1 saddle, $-x_1^2+x_2^2+x_3^2+x_4^2$.

2-handles attach to the union of 0- and 1-handles and correspond to an index = 2 saddle $-x_1^2-x_2^2+x_3^2+x_4^2$.

3-handles attach to the previous union of 0-, 1-, and 2-handles and correspond to an index = 3 saddles, $-x_1^2-x_2^2-x_3^2+x_4^2$.

4-handles correspond to a local maxima, $-x_1^2-x_2^2-x_3^2-x_4^2$.

An interesting aspect of handle bodies is that there are moves which slide one $k$-handle $h_1$, over a second $k$-handle $h_2$, which change the attaching maps, but do not change the diffeomorphism type of the manifold being described. The geometric operation of sliding one 2-handle over another has an algebraic analog in the diagrammatic formalism of TQFTs and UMTCs. First, we explain the geometric move and then the analog.

%%%%%%%%%%%%%%%%%%%%%%%%%%%%%%%%%%%%%%%%%%%%%%%%%%%%%%%%%%%%%%%%%%%%%%%%%%%%%%%%%%%%%%%%%%
\begin{figure}[t!]
  \labellist
  % labels for (a)
  \pinlabel $\text{slide}$ at 530 1700
  \pinlabel $\text{here}$ at 530 1630
  \pinlabel $\text{attach }h_1$ at 50 1000
  \pinlabel $\text{attach }h_2$ at 850 900
  \pinlabel \Huge{$\leadsto$} at 1400 1300
  \pinlabel $\text{(a)}$ at 530 800
  % labels for (b)
  \pinlabel $\text{attach }h_1^{\prime}$ at 1550 1000
  \pinlabel $\text{attach }h_2$ at 2150 900
  \pinlabel $\text{(b)}$ at 1950 800
  % labels for (c)
  \pinlabel $h_1$ at 500 300
  \pinlabel $h_2$ at 980 300
  \pinlabel \Huge{$\leadsto$} at 1350 200
  \pinlabel $\text{(c)}$ at 700 -100
  % labels for (d)
  \pinlabel $h_1^{\prime}$ at 1720 600
  \pinlabel $h_2$ at 2170 300
  \pinlabel $\text{(d)}$ at 2040 -100
  \endlabellist
  \begin{center}
  \includegraphics[width=0.9\textwidth]{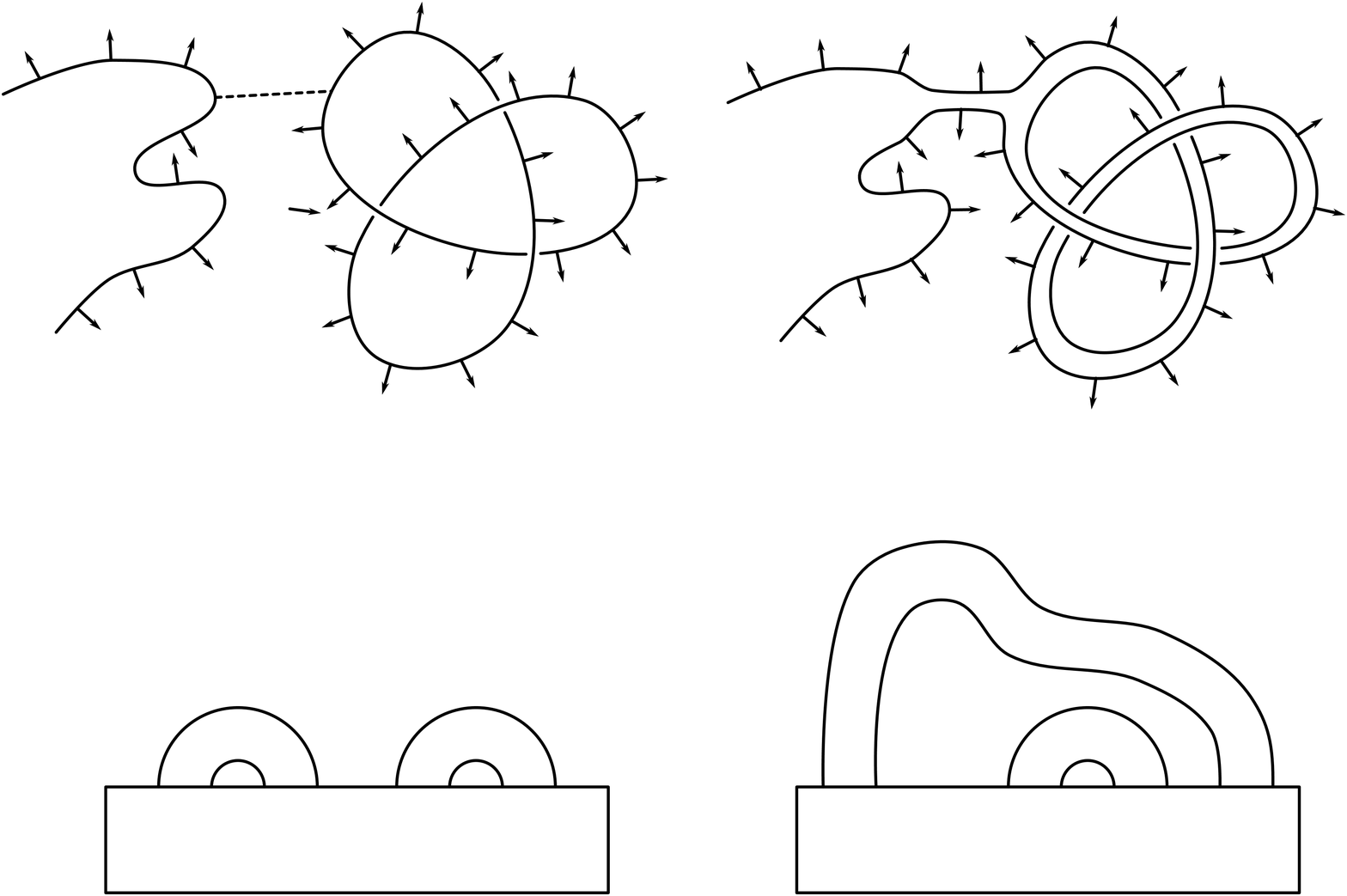}
  \vspace{0.5cm}
  \caption{Sliding handles.}
  \label{fig:handle_slide}
  \vspace{0.5cm}
  \end{center}
\end{figure}
%%%%%%%%%%%%%%%%%%%%%%%%%%%%%%%%%%%%%%%%%%%%%%%%%%%%%%%%%%%%%%%%%%%%%%%%%%%%%%%%%%%%%%%%%%%%%

Passing a 2-handle $h_1$ over another $h_2$ means transforming the two solid tori attaching regions, drawn as framed loops in a 3-manifold, as shown in Fig.~\ref{fig:handle_slide}(a) and (b). The framing describes how $\partial B^2\times B^2$ is identified or ``glued'' to a neighborhood of the loop. An idea of how the 4D-handles are sliding is given by the sketch in Fig.~\ref{fig:handle_slide}(c) and (d), in which the dimensions have been cut in half.

As far as the effect on the boundary 3-manifold is concerned, the attachment of a 2-handle realizes a surgery (sometimes called Dehn surgery), meaning that the solid torus to which the attaching region is glued is removed and then another replacement solid torus, in this case $B^2\times\partial B^2$, is glued back in. The meridional loop, $\partial B^2\times\ast$ of the new solid torus, matches with whichever longitude on the original solid torus is dictated by the framing vector. From this point of view, the rules (Table~\ref{tab:attaching_region}) for sliding handles amounts to a way of recognizing that surgery on two different framed links yield the same 3-manifold, after surgery. The subject which decides when two framed links yield (upon surgery) the same 3-manifold is often called ``Kirby calculus.''

The following diagrammatic calculation
\begin{eqnarray}
&&
\pspicture[shift=-0.7](-1.6,-0.3)(1.5,1.3)
\small
  \psset{linewidth=0.9pt,linecolor=black,arrowscale=1.5,arrowinset=0.15}
  \psellipse[border=1.5pt](0.4,0.5)(0.8,0.35)
  \psline{->}(0.2,0.17)(0.3,0.15)
  \rput[bl]{0}(-0.1,-0.2){$\omega_0$}
  \psline(-1,0)(-1,1)
  \rput[bl]{0}(-1.35,0.5){$a$}
  \psline{->}(-1,0.5)(-1,0.75)
\endpspicture
= \sum_{b} \frac{d_b}{ \mathcal{D}^2}
\pspicture[shift=-0.7](-1.6,-0.3)(1.5,1.3)
\small
  \psset{linewidth=0.9pt,linecolor=black,arrowscale=1.5,arrowinset=0.15}
  \psellipse[border=1.5pt](0.4,0.5)(0.8,0.35)
  \psline{->}(0.2,0.17)(0.3,0.15)
  \rput[bl]{0}(-0.1,-0.2){$b$}
  \psline(-1,0)(-1,1)
  \rput[bl]{0}(-1.35,0.5){$a$}
  \psline{->}(-1,0.5)(-1,0.75)
\endpspicture
\notag \\
&& \quad
=
\sum_{b,c,\mu} \frac{d_b}{ \mathcal{D}^2} \sqrt{\frac{d_c}{ d_a d_b}}
\pspicture[shift=-1.1](-1.6,-0.8)(1.6,1.4)
\small
  \psset{linewidth=0.9pt,linecolor=black,arrowscale=1.5,arrowinset=0.15}
  \psellipticarc[border=1.5pt](0.4,0.5)(1.1,0.6){-160}{160}
  \psline(-0.485,0.825)(-1,1.1)
  \psline(-0.485,0.175)(-1,-0.1)
  \psline{->}(-0.485,0.825)(-0.87125,1.03125)
  \rput[bl]{0}(-1.3,1.05){$a$}
  \psline{-<}(-0.485,0.175)(-0.87125,-0.03125)
  \rput[bl]{0}(-1.3,-0.25){$a$}
  \psline(-0.485,0.175)(-0.485,0.825)
  \psline{->}(-0.485,0.175)(-0.485,0.62)
  \rput[bl]{0}(-0.8,0.4){$c$}
  \psline{->}(0.2,-0.08)(0.3,-0.1)
  \rput[bl]{0}(0.07,-0.5){$b$}
  \scriptsize
  \rput[bl]{0}(-0.4,0.65){$\mu$}
  \rput[bl]{0}(-0.4,0.15){$\mu$}
\endpspicture
=
\sum_{b,c,\mu} \frac{d_b}{ \mathcal{D}^2} \sqrt{\frac{d_c}{ d_a d_b}}
\pspicture[shift=-1.1](-1.6,-0.8)(1.5,1.4)
\small
  \psset{linewidth=0.9pt,linecolor=black,arrowscale=1.5,arrowinset=0.15}
  \psellipticarc[border=1.5pt](0.4,0.5)(0.8,0.35){20}{340}
  \psline{-<}(0.2,0.17)(0.3,0.15)
  \rput[bl]{0}(0.1,0.31){$c$}
  \psellipticarc(0.4,0.5)(1.1,0.6){-160}{-20}
  \psellipticarc(0.4,0.5)(1.1,0.6){20}{160}
  \psline(1.285,0.825)(0.98,0.72)
  \psline(1.285,0.175)(0.98,0.28)
  \psline(-0.485,0.825)(-1,1.1)
  \psline(-0.485,0.175)(-1,-0.1)
  \psline{->}(-0.485,0.825)(-0.87125,1.03125)
  \rput[bl]{0}(-1.3,1.05){$a$}
  \psline{-<}(-0.485,0.175)(-0.87125,-0.03125)
  \rput[bl]{0}(-1.3,-0.25){$a$}
  \psline(0.98,0.28)(0.98,0.72)
  \psline{->}(0.98,0.28)(0.98,0.62)
  \rput[bl]{0}(1.15,0.35){$b$}
  \scriptsize
  \rput[bl]{0}(0.7,0.55){$\mu$}
  \rput[bl]{0}(0.7,0.25){$\mu$}
\endpspicture
\notag \\
&&
\quad =
\sum_{b,c,\mu} \frac{d_b}{ \mathcal{D}^2 d_a}
\pspicture[shift=-0.7](-1.6,-0.3)(1.6,1.3)
\small
  \psset{linewidth=0.9pt,linecolor=black,arrowscale=1.5,arrowinset=0.15}
  \psellipse[border=1.5pt](0.4,0.5)(0.8,0.35)
  \psline{-<}(0.2,0.17)(0.3,0.15)
  \rput[bl]{0}(0.1,0.31){$c$}
  \psellipticarc(0.4,0.5)(1.1,0.6){-160}{160}
  \psline(-0.485,0.825)(-1,1.1)
  \psline(-0.485,0.175)(-1,-0.1)
  \psline{->}(-0.485,0.825)(-0.87125,1.03125)
  \rput[bl]{0}(-1.3,1.05){$a$}
\endpspicture
=
\pspicture[shift=-0.7](-1.6,-0.3)(1.5,1.3)
\small
  \psset{linewidth=0.9pt,linecolor=black,arrowscale=1.5,arrowinset=0.15}
  \psellipse[border=1.5pt](0.4,0.5)(0.8,0.35)
  \psline{->}(0.2,0.17)(0.3,0.15)
  \rput[bl]{0}(-0.05,0.3){$\omega_0$}
  \psellipticarc(0.4,0.5)(1.1,0.6){-160}{160}
  \psline(-0.485,0.825)(-1,1.1)
  \psline(-0.485,0.175)(-1,-0.1)
  \psline{->}(-0.485,0.825)(-0.87125,1.03125)
  \rput[bl]{0}(-1.3,1.05){$a$}
\endpspicture
\label{eq:handle_slide}
\end{eqnarray}
establishes the handle slide property for $\omega_0$-loops. This shows that, within the UMTC formalism, if a framed loop $\gamma_2$ is an $\omega_0$-loop, then the partition function $Z$ is unaffected by sliding an arbitrarily labeled loop $\gamma_1$ over $\gamma_2$. For simplicity, in Eq.~(\ref{eq:handle_slide}), we have shown only an arc segment of $\gamma_1$ (labeled with charge $a$) and $\gamma_2$ as an ellipse, but one may think of $\gamma_2$ as a knot, as in Fig.~\ref{fig:handle_slide}. Thus, a loop labeled by $\omega_0$ has the same handle slide property as a 2-handle $h_2$. This justifies interpreting $\omega_0$-labeled framed loops in all diagrams of states or density matrices as being ``surgered.'' That is, the diagram effectively exists in a topologically exotic space-time 3-manifold created by surgery on the $\omega_0$-loops, and therefore consists only of the loops not labeled by $\omega_0$.

There is an immediate generalization from $\omega_0$-loops to $\omega_a$-loops. After doing the surgery indicated by $\omega_0$, the loop labeled by $a$ slides into a copy of the core $0\times\partial B^2\subset B^2\times\partial B^2$ of the replacement solid torus (with product normal framing). Thus, any loop labeled by $\omega_{a}$ may also be interpreted as surgered out in the effective diagram, but with the difference that there will now be a Wilson loop with charge $a$ (and product framing) running along the core of the replacement solid torus. This is represented diagrammatically by
\begin{equation}
\pspicture[shift=-0.55](-0.4,-0.3)(1.3,1.3)
\small
  \psset{linewidth=0.9pt,linecolor=black,arrowscale=1.5,arrowinset=0.15}
  \psellipse[linewidth=0.9pt,linecolor=black,border=0](0.4,0.5)(0.8,0.35)
  \psset{linewidth=0.9pt,linecolor=black,arrowscale=1.4,arrowinset=0.15}
  \psline{->}(0.2,0.17)(0.3,0.15)
  \rput[bl]{0}(-0.1,-0.2){$\omega_a$}
  \endpspicture
=
\pspicture[shift=-0.55](-0.4,-0.3)(1.3,1.3)
\small
  \psset{linewidth=0.9pt,linecolor=black,arrowscale=1.5,arrowinset=0.15}
  \psarc[linewidth=0.9pt,linecolor=black,border=0pt] (1.15,0.5){0.25}{0}{360}
  \psellipse[linewidth=0.9pt,linecolor=black,border=1.5pt](0.4,0.5)(0.8,0.35)
  \psset{linewidth=0.9pt,linecolor=black,arrowscale=1.4,arrowinset=0.15}
  \psline{->}(0.2,0.17)(0.3,0.15)
  \rput[bl]{0}(-0.1,-0.2){$\omega_0$}
 \psarc[linewidth=0.9pt,linecolor=black,border=1.5pt] (1.15,0.5){0.25}{90}{180}
 \psline{->}(1.398,0.56)(1.399,0.58)
   \rput[bl]{0}(1.5,0.4){$a$}
  \endpspicture
\end{equation}
Similarly, one can formally sum over the charge values $a$ of $\omega_a$-loops in such diagrams.

In general, curves labeled by $\omega_{a}$ do not have a particularly convenient handle slide property. However, there is a nice identity for sliding an $\omega_{a}$-loop over an $\omega_{b}$-loop when $b$ is an Abelian anyon:
\begin{eqnarray}
&&
\pspicture[shift=-0.7](-0.7,-0.3)(4.1,1.3)
\small
  \psset{linewidth=0.9pt,linecolor=black,arrowscale=1.5,arrowinset=0.15}
  \psellipse[border=1.5pt](0.4,0.5)(0.8,0.35)
  \psline{->}(0.2,0.17)(0.3,0.15)
  \rput[bl]{0}(-0.1,-0.2){$\omega_a$}
  \psellipse[border=1.5pt](3,0.5)(0.8,0.35)
  \psline{->}(2.8,0.17)(2.9,0.15)
  \rput[bl]{0}(2.5,-0.2){$\omega_b$}
\endpspicture
=
\pspicture[shift=-0.7](-0.7,-0.3)(4.1,1.3)
\small
  \psset{linewidth=0.9pt,linecolor=black,arrowscale=1.5,arrowinset=0.15}
  \psellipse[border=0](0.4,0.5)(0.8,0.35)
  \psline{->}(0.2,0.17)(0.3,0.15)
  \rput[bl]{0}(-0.1,-0.2){$\omega_a$}
  \psarc[border=0pt] (3.75,0.5){0.25}{0}{360}
  \psellipse[border=1.5pt](3,0.5)(0.8,0.35)
  \psline{->}(2.8,0.17)(2.9,0.15)
  \rput[bl]{0}(2.5,-0.2){$\omega_0$}
  \psarc[border=1.5pt] (3.75,0.5){0.25}{90}{180}
  \psline{->}(3.998,0.56)(3.999,0.58)
  \rput[bl]{0}(4.1,0.4){$b$}
\endpspicture
\notag \\
&&
\quad =
\pspicture[shift=-0.7](-0.7,-0.3)(5,1.3)
\small
  \psset{linewidth=0.9pt,linecolor=black,arrowscale=1.5,arrowinset=0.15}
  \psarc[border=0pt] (4,0.5){0.5}{0}{360}
  \psellipticarc[border=1.5pt](0.4,0.5)(0.8,0.35){10}{350}
  \psline{->}(0.2,0.17)(0.3,0.15)
  \rput[bl]{0}(-0.1,-0.2){$\omega_a$}
  \psellipse[border=1.5pt](3,0.5)(0.8,0.35)
  \psline{->}(2.8,0.17)(2.9,0.15)
  \rput[bl]{0}(2.55,0.275){$\omega_0$}
  \psellipticarc[border=1.5pt](3,0.5)(1.1,0.6){-173.5}{173.5}
  \psline(1.115,0.635)(1.95,0.635)
  \psline(1.115,0.365)(1.95,0.365)
  \psarc[border=1.5pt] (4,0.5){0.5}{90}{180}
  \psline{->}(4.5,0.56)(4.501,0.58)
  \rput[bl]{0}(4.6,0.4){$b$}
\endpspicture
=
\pspicture[shift=-0.7](-0.7,-0.3)(5,1.3)
\small
  \psset{linewidth=0.9pt,linecolor=black,arrowscale=1.5,arrowinset=0.15}
  \psellipticarc[border=1.5pt](0.4,0.5)(0.8,0.35){10}{350}
  \psline{->}(0.2,0.17)(0.3,0.15)
  \rput[bl]{0}(-0.1,-0.2){$\omega_a$}
  \psarc[border=0pt] (3.75,0.5){0.25}{0}{360}
  \psellipse[border=1.5pt](3,0.5)(0.8,0.35)
  \psline{->}(2.8,0.17)(2.9,0.15)
  \rput[bl]{0}(2.55,0.275){$\omega_0$}
  \psarc[border=1.5pt] (3.75,0.5){0.25}{90}{180}
  \psline{->}(3.998,0.56)(3.999,0.58)
  \rput[bl]{0}(4.1,0.4){$b$}
  \psarc[border=0pt] (4.625,0.5){0.25}{0}{360}
  \psellipticarc[border=1.5pt](3.2875,0.5)(1.3875,0.6){-174.5}{174.5}
  \psarc[border=1.5pt] (4.625,0.5){0.25}{90}{180}
  \psline{->}(4.873,0.56)(4.874,0.58)
  \rput[bl]{0}(4.975,0.4){$b$}
  \psline(1.115,0.635)(1.96,0.635)
  \psline(1.115,0.365)(1.96,0.365)
\endpspicture
\notag \\
&&
\quad =
\pspicture[shift=-0.7](-0.7,-0.3)(4.1,1.3)
\small
  \psset{linewidth=0.9pt,linecolor=black,arrowscale=1.5,arrowinset=0.15}
  \psellipticarc[border=1.5pt](0.4,0.5)(0.8,0.35){10}{350}
  \psline{->}(0.2,0.17)(0.3,0.15)
  \rput[bl]{0}(-0.1,-0.2){$\omega_{a\times b}$}
  \psellipse[border=1.5pt](3,0.5)(0.8,0.35)
  \psline{->}(2.8,0.17)(2.9,0.15)
  \rput[bl]{0}(2.55,0.275){$\omega_b$}
  \psellipticarc(3,0.5)(1.1,0.6){-173.5}{173.5}
  \psline(1.115,0.635)(1.95,0.635)
  \psline(1.115,0.365)(1.95,0.365)
\endpspicture
\label{id:loop_slide}
\end{eqnarray}
This identity will play a key role in simplifying the analysis of both twisted and untwisted interferometers in Ising-type systems, as it allows us to slide $\omega_{a}$-loops over $\omega_{b}$-loops when $b=\psi$ is the (Abelian) fermion charge of the Ising theory.

Using the handle slide property of $\omega_0$-loops, the post-measurement density matrix of Eq.~(\ref{eq:target_projected_omega}) can be rewritten (as previously mentioned) as
\begin{equation}
{\rho}_{a}^{AC} = \frac{1}{{\Pr}_{AC}(a)}
\pspicture[shift=-2.9](-2.8,-3)(2.8,2.5)
  \small
%%%%% Box:
  \psline(-2.5,-0.5)(-2.5,0.5)
  \psline(-2.5,-0.5)(2.5,-0.5)
  \psline(-2.5,0.5)(2.5,0.5)
  \psline(2.5,-0.5)(2.5,0.5)
%%%%% Line connections:
  \psset{linewidth=0.9pt,linecolor=black,arrowscale=1.5,arrowinset=0.15}
  \psline(2.0,0.5)(2.0,2)
  \psline(0.0,0.5)(0.0,2)
  \psline(2.0,-0.5)(2.0,-2)
  \psline(0.0,-0.5)(0.0,-2)
  \psline(-2.0,0.5)(-2.0,2)
  \psline(-2.0,-0.5)(-2.0,-2)
%%%%% Arrows:
  \psline{->}(2.0,1.5)(2.0,1.75)
  \psline{-<}(2.0,-1.5)(2.0,-1.75)
  \psline{->}(-2.0,1.5)(-2.0,1.75)
  \psline{-<}(-2.0,-1.5)(-2.0,-1.75)
%%%%% Leg Labels:
  \rput[bl](1.9,2.05){$C_1$}
  \rput[bl](-0.1,2.1){$A$}
  \rput[bl](1.9,-2.45){$C_1^{\prime}$}
  \rput[bl](-0.1,-2.35){$A^{\prime}$}
  \rput[bl](-2.3,2.05){$C_2$}
  \rput[bl](-2.3,-2.45){$C_2^{\prime}$}
%%%% Probe Loop:
  \psellipse[linewidth=0.9pt,linecolor=black,border=0.1](0.0,1.1)(1.0,0.3)
   \psline{<-}(-0.2,0.83)(-0.25,0.83)
  \psline[linewidth=0.9pt,linecolor=black,border=0.1](0.0,1.2)(0.0,1.8)
  \psellipse[linewidth=0.9pt,linecolor=black,border=0.1](0.0,-1.1)(1.0,0.3)
   \psline{<-}(-0.2,-1.37)(-0.25,-1.37)
  \psline[linewidth=0.9pt,linecolor=black,border=0.1](0.0,-0.6)(0.0,-1.2)
   \psline{-<}(0.0,-1.5)(0.0,-1.75)
  \psline{->}(0.0,1.5)(0.0,1.75)
  \psellipse[linewidth=0.9pt,linecolor=black,border=0.05](1.5,0.0)(0.3,1.5)
  \psline{->}(1.215,-0.13)(1.215,-0.1)
  \psframe[linewidth=0.9pt,linecolor=white,border=0.1,fillcolor=white,fillstyle=solid](1.7,-0.5)(1.8,0.5)
  \psline(2.0,0.5)(1.5,0.5)
  \psline(2.0,-0.5)(1.5,-0.5)
  \psellipse[linewidth=0.9pt,linecolor=black,border=0.05](-1.5,0.0)(-0.3,1.5)
  \psline{->}(-1.19,-0.13)(-1.19,-0.1)
  \psframe[linewidth=0.9pt,linecolor=white,border=0.1,fillcolor=white,fillstyle=solid](-1.7,-0.5)(-1.8,0.5)
  \psline(-2.0,0.5)(-1.5,0.5)
  \psline(-2.0,-0.5)(-1.5,-0.5)
%%%%% Labels:
  \rput[bl]{0}(-0.3,-0.15){$\rho^{AC}$}
  \scriptsize
  \rput[bl](1.35,-0.3){$\omega_{0}$}
  \rput[bl](-0.8,0.6){$\omega_{a}$}
  \rput[bl](-1.65,-0.3){$\omega_{0}$}
  \rput[bl](-0.8,-1.6){$\omega_{a}$}
 \endpspicture
\label{eq:target_projected_omega_slid}
.
\end{equation}
Note that the use of two $\omega_a$-loops here is redundant, since one can move one of them to the other one's position using handle slide and $\omega$-loops are idempotent (i.e. they are projectors).

\subsection{The Effective Surgical Description of Interferometry}
\label{sec:surgical_description}

The density matrix formalism replaces a state vector $\ket{\psi}$ with a state operator $\rho$, equal to $\rho = \ket{\psi}\bra{\psi}$ for a pure state. Similarly in the density matrix formalism, a space-time evolution in $\text{Hom}(W,V)$ carrying an initial $\psi_0$ to $\psi_1$ becomes an operator in $V^{\ast}\otimes V$ by forming $\text{Hom}(W,V)\otimes\text{Hom}^{\ast}(W,V)$ and tracing out $W$. Topologically, the density matrix components are (superpositions of) diagrams in a space-time glued to a copy of itself reflected across a time $t=0$ plane. The diagrams in Eqs.~(\ref{eq:target_projected_omega}) and (\ref{eq:target_projected_omega_slid}) should be interpreted in this way.

In topological language, the conclusion of Refs.~\cite{Bonderson07b,Bonderson07c,Bonderson13b}, as recapitulated in Section~\ref{sec:what_it_does} is that (up to exponentially suppressed corrections) the effective diagram for the partition function is the probabilistic combination of Dehn surgeries and Wilson loops indicated in Fig.~\ref{fig:handle_bodies}. Note that, while the exact partition function is given by $4^N$ terms with probe anyon Wilson loops, the effective diagram has no probe anyons in it. It only has a small number of Dehn surgeries, some with Wilson loops at the core. Surgeries on $\omega_0$-loops are ``ordinary'' and for the $\omega_a$-loop surgeries, one input is a probabilistically determined charge $a$ along the core circle (Wilson loop) of the replacement solid torus. This may also involve a sum of simple charges $\omega_{\mathcal{A}} = \sum_{a \in \mathcal{A}} \omega_a$, if one wishes to treat the case where the probe anyons do not distinguish all topological charge types, i.e. $M_{a,b}=M_{a^{\prime},b}$ for all $a, a^{\prime} \in \mathcal{A}$. In this case, the Wilson loop has a superposition of charges $a\in \mathcal {A}$, i.e. is treated as a formal linear combination of diagrams.

%%%%%%%%%%%%%%%%%%%%%%%%%%%%%%%%%%%%%%%%%%%%%%%%%%%%%%%%%%%%%%%%%%%%%%%%%%%%%%%%%%%%%%%%%%%%%%%%
\begin{figure}[t!]
  \labellist
  \pinlabel $\text{doubled}$ at -60 140
  \pinlabel $\text{space-time }X$ at -60 120
  \pinlabel $V^{\ast}$ at 260 230
  \pinlabel $\otimes$ at 260 190
  \pinlabel $W$ at 260 150
  \pinlabel $\otimes$ at 260 120
  \pinlabel $W^{\ast}$ at 260 85
  \pinlabel $\otimes$ at 260 50
  \pinlabel $V$ at 260 10
  \pinlabel $\rightarrow V\otimes V^{\ast}$ at 350 120
  \endlabellist
  \begin{center}
  \includegraphics[width=1.5in]{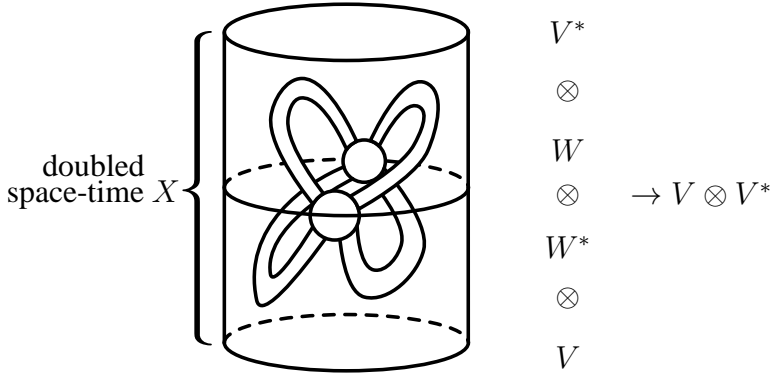}
  \hspace{5cm}
  \caption{Density matrix as a diagram in space-time glued to a reflected copy of itself.}
  \label{fig:double_diagram}
  \end{center}
\end{figure}

\subsection{Twisted Interferometers}
\label{sec:twisted_interferometers}

%%%%%%%%%%%%%%%%%%%%%%%%%%%%%%%%%%%%%%%%%%%%%%%%%%%%%%%%%%%%%%%%%%%%%%%%%%%%%%%%%%%%%%%%%%%%%%%%%%%%%%%%%%%%%
\begin{figure}[t!]
\begin{center}
  \includegraphics[scale=0.4]{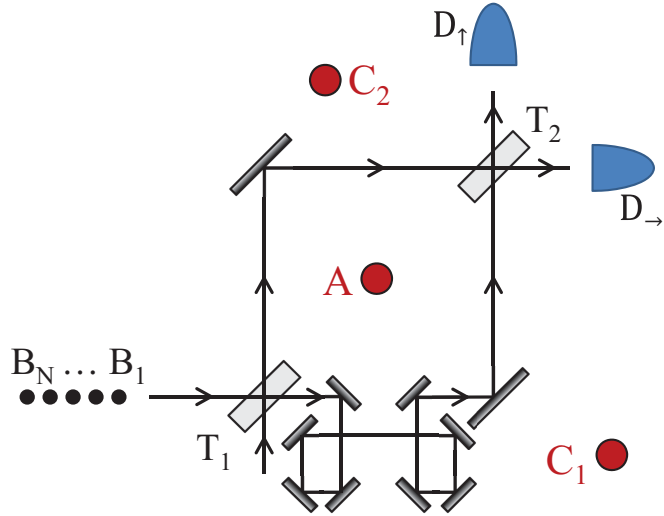}
  \caption{An idealized Mach-Zehnder anyonic interferometer with a doubly twisted path in its right arm.}
  \label{fig:Double_Twisted_int}
\end{center}
\end{figure}
%%%%%%%%%%%%%%%%%%%%%%%%%%%%%%%%%%%%%%%%%%%%%%%%%%%%%%%%%%%%%%%%%%%%%%%%%%%%%%%%%%%%%%%%%%%%%%%%%%%%%%%%%%%%%

Now that we have established the topological language, the modification necessary to compute the effect on the partition function $Z$ of twisting the arms amounts to cutting the handle-body $H$ out of the doubled space-time and gluing back in with certain twists.

\subsection{Computing the Consequence of Twisting}
\label{sec:twist_consequence}

%%%%%%%%%%%%%%%%%%%%%%%%%%%%%%%%%%%%%%%%%%%%%%%%%%%%%%%%%%%%%%%%%%%%%%%%%%%%%
\begin{figure}[t!]
  \labellist
  % labels for (a)
  \pinlabel $\omega_a$ at 50 500
  \pinlabel $\omega_a$ at -50 450
  \pinlabel $\omega_0$ at 720 320
  \pinlabel $\text{(a)}$ at 350 -110
  % labels for (b)
  \pinlabel $\omega_a$ at 1400 700
  \pinlabel $\omega_0$ at 900 330
  \pinlabel $\omega_0$ at 1830 330
  \pinlabel $\omega_a$ at 1370 -20
  \pinlabel $\text{(b)}$ at 1370 -110
  \endlabellist
\begin{center}
\includegraphics[width=0.9\textwidth]{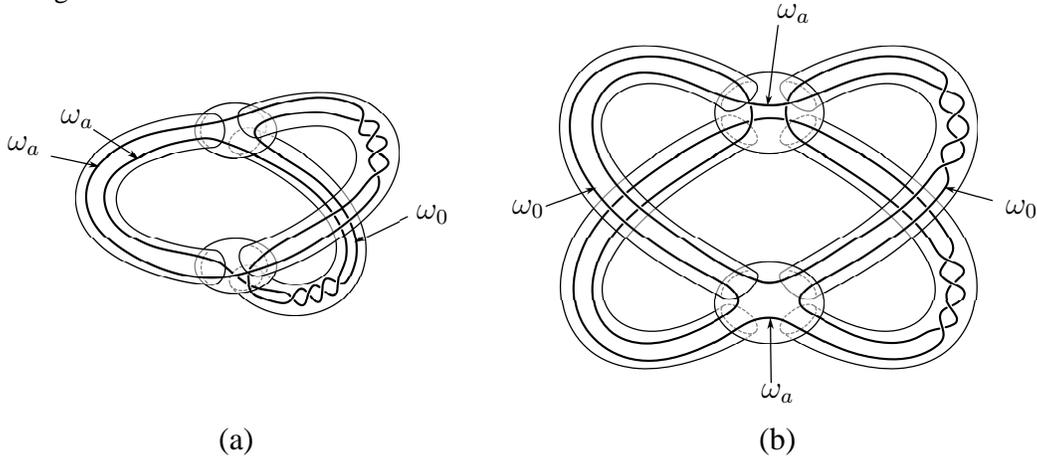}
  \vspace{0.75cm}
  \caption{(a) $H\subset X$ and (b) $H^{\prime}\subset X$ re-glued back into the doubled space-time $X$ after introducing twists into the handles. Here we show the twisting, $+2$ Dehn twists applied to $\gamma$ and $-2$ Dehn twists applied to $\bar{\gamma}$, corresponding to a double twist implemented in the right arm of the interferometer.}
  \label{fig:repositioned}
\end{center}
\end{figure}
%%%%%%%%%%%%%%%%%%%%%%%%%%%%%%%%%%%%%%%%%%%%%%%%%%%%%%%%%%%%%%%%%%%%%%%%%%%%%

The operation of an idealized anyonic interferometer is described by a few (generalized) surgeries within the handle body $H$ or $H^{\prime}$ inside the doubled space-time manifold $X$, as shown in Fig.~\ref{fig:handle_bodies}. In this surgery formulation, introducing probe anyon twisting into the arms of the interferometer is accounted for by removing the handle body $H$ or $H^{\prime}$ from the doubled space-time and then re-gluing it back into $X\setminus H$ or $X\setminus H^{\prime}$, respectively, with additional twists as shown in Fig.~\ref{fig:repositioned}.

Let $l$ and $r$ represent the number of full twists imposed on the left and right arms, respectively. The appropriate re-gluing of $H$ or $H^{\prime}$ is induced by a number of Dehn twists applied to the loops $\beta, \bar{\beta}, \gamma, \bar{\gamma}$ in Fig.~\ref{fig:handle_bodies} according to the rules
\begin{equation}
  \begin{tabular}{c|c}
    loop & \# of Dehn twists\\\hline
    $\gamma$ & $r$\\
    $\bar{\gamma}$ & $-r$\\
    $\beta$ & $l$\\
    $\bar{\beta}$ & $-l$
  \end{tabular}
\label{eq:twist_rules}
\end{equation}

The effect of opposite (mirror image) twisting leaves the framing of the $\omega_0$-labeled curves unchanged. In the re-glued $H$ or $H^{\prime}$, the $\omega_a$-loops and $\omega_0$-loops are repositioned as shown in Fig.~\ref{fig:repositioned} for $r=2$ and $l=0$, i.e. an interferometer with a double twist in its right arm.

Thus, the conclusion of our topological/diagrammatic analysis is:

Using the computational rules inherent in the definition of a (2+1)D TQFT (i.e. UMTC), the effective result of $(l,r)$-twisted anyonic interferometers (ignoring exponentially suppressed corrections, multiple passes, and probe-probe interactions) by inserting the Wilson loops, as shown in Fig.~\ref{fig:repositioned} for $(l,r) = (0,2)$, as in Fig.~\ref{fig:handle_bodies} with Dehn twists applied to the loops $\gamma$, $\bar{\gamma}$, $\beta$, and $\bar{\beta}$ according to the rules in Eq.~(\ref{eq:twist_rules}) and evaluating the density matrix $Z$. Diagrammatically, this can be represented by
\begin{equation}
\tilde{\rho}_{a}^{AC} = \frac{1}{\tilde{\Pr}_{AC}(a)}
\pspicture[shift=-3.5](-2.8,-3.6)(2.8,3.5)
  \small
%%%%% Box:
  \psline(-2.5,-0.5)(-2.5,0.5)
  \psline(-2.5,-0.5)(2.5,-0.5)
  \psline(-2.5,0.5)(2.5,0.5)
  \psline(2.5,-0.5)(2.5,0.5)
%%%%% Line connections:
  \psset{linewidth=0.9pt,linecolor=black,arrowscale=1.5,arrowinset=0.15}
  \psline(2.0,0.5)(2.0,3.0)
  \psline(0.0,0.5)(0.0,3.0)
  \psline(2.0,-0.5)(2.0,-3.0)
  \psline(0.0,-0.5)(0.0,-3.0)
  \psline(-2.0,0.5)(-2.0,3.0)
  \psline(-2.0,-0.5)(-2.0,-3.0)
%%%%% Arrows:
  \psline{->}(2.0,1.5)(2.0,2.75)
  \psline{->}(0.0,1.5)(0.0,2.75)
  \psline{-<}(2.0,-1.5)(2.0,-2.75)
  \psline{-<}(0.0,-1.5)(0.0,-2.75)
  \psline{->}(-2.0,1.5)(-2.0,2.75)
  \psline{-<}(-2.0,-1.5)(-2.0,-2.75)
%%%%% Leg Labels:
  \rput[bl](1.9,3.05){$C_1$}
  \rput[bl](-0.1,3.1){$A$}
  \rput[bl](1.9,-3.45){$C_1^{\prime}$}
  \rput[bl](-0.1,-3.35){$A^{\prime}$}
  \rput[bl](-2.3,3.05){$C_2$}
  \rput[bl](-2.3,-3.45){$C_2^{\prime}$}
%%%% tau Loops:
    \psarc[linewidth=0.9pt,linecolor=black]{<-}(-1.0,1.5){0.6}{80}{450}
    \psarc[linewidth=0.9pt,linecolor=black]{<-}(1.0,1.5){0.6}{80}{450}
    \psarc[linewidth=0.9pt,linecolor=black]{<-}(-1.0,-1.5){0.6}{-100}{270}
    \psarc[linewidth=0.9pt,linecolor=black]{<-}(1.0,-1.5){0.6}{-100}{270}
%%%% Probe Loop:
  \psellipse[linewidth=0.9pt,linecolor=black,border=0.05](0.0,1.5)(1.0,0.3)
  \psellipse[linewidth=0.9pt,linecolor=black,border=0.05](0.0,-1.5)(1.0,0.3)
 \psline{<-}(-0.1,1.23)(-0.15,1.23)
 \psline{<-}(-0.1,-1.78)(-0.15,-1.78)
  \psline[linewidth=0.9pt,linecolor=black,border=0.1](0.0,1.5)(0.0,2.0)
  \psline[linewidth=0.9pt,linecolor=black,border=0.1](0.0,-0.8)(0.0,-1.5)
  \psellipse[linewidth=0.9pt,linecolor=black,border=0.05](1.5,0.0)(0.3,1.5)
  \psline{->}(1.215,-0.13)(1.215,-0.1)
  \psframe[linewidth=0.9pt,linecolor=white,border=0.1,fillcolor=white,fillstyle=solid](1.7,-0.5)(1.8,0.5)
  \psline(2.0,0.5)(1.5,0.5)
  \psline(2.0,-0.5)(1.5,-0.5)
  \psellipse[linewidth=0.9pt,linecolor=black,border=0.05](-1.5,0.0)(-0.3,1.5)
  \psline{->}(-1.19,-0.13)(-1.19,-0.1)
  \psframe[linewidth=0.9pt,linecolor=white,border=0.1,fillcolor=white,fillstyle=solid](-1.7,-0.5)(-1.8,0.5)
  \psline(-2.0,0.5)(-1.5,0.5)
  \psline(-2.0,-0.5)(-1.5,-0.5)
    \psarc[linewidth=0.9pt,linecolor=black,border=0.05](-1.0,1.5){0.6}{0}{45}
    \psarc[linewidth=0.9pt,linecolor=black,border=0.05](-1.0,1.5){0.6}{135}{225}
    \psarc[linewidth=0.9pt,linecolor=black,border=0.05](1.0,1.5){0.6}{-45}{45}
    \psarc[linewidth=0.9pt,linecolor=black,border=0.05](1.0,1.5){0.6}{135}{180}
    \psarc[linewidth=0.9pt,linecolor=black,border=0.05](-1.0,-1.5){0.6}{0}{90}
    \psarc[linewidth=0.9pt,linecolor=black,border=0.05](-1.0,-1.5){0.6}{135}{225}
    \psarc[linewidth=0.9pt,linecolor=black,border=0.05](1.0,-1.5){0.6}{-45}{45}
    \psarc[linewidth=0.9pt,linecolor=black,border=0.05](1.0,-1.5){0.6}{90}{180}
%%%%% Labels:
  \rput[bl]{0}(-0.3,-0.15){$\rho^{AC}$}
  \scriptsize
  \rput[bl](1.35,-0.3){$\omega_{0}$}
  \rput[bl](-0.4,0.9){$\omega_{a}$}
  \rput[bl](-1.6,-0.3){$\omega_{0}$}
  \rput[bl](-0.4,-2.1){$\omega_{a}$}
  \rput[bl](0.8,2.25){$\tau^{r}$}
  \rput[bl](-1.2,2.25){$\tau^{l}$}
  \rput[bl](0.75,-2.5){$\tau^{-r}$}
  \rput[bl](-1.25,-2.5){$\tau^{-l}$}
 \endpspicture
\label{eq:twisted_target_projected_omega}
\end{equation}
where a $\tau^m$-loop, given by
\begin{equation}
\pspicture[shift=-0.55](-0.25,-0.1)(0.9,1.3)
\small
  \psset{linewidth=0.9pt,linecolor=black,arrowscale=1.5,arrowinset=0.15}
  \psellipse[linewidth=0.9pt,linecolor=black,border=0](0.4,0.5)(0.4,0.18)
  \psset{linewidth=0.9pt,linecolor=black,arrowscale=1.4,arrowinset=0.15}
  \psline{->}(0.2,0.37)(0.3,0.34)
  \rput[bl]{0}(-0.4,0.0){$\tau^{m}$}
\endpspicture
= \sum_{a} \theta_a^{m}
\pspicture[shift=-0.55](-0.25,-0.1)(0.9,1.3)
\small
  \psset{linewidth=0.9pt,linecolor=black,arrowscale=1.5,arrowinset=0.15}
  \psellipse[linewidth=0.9pt,linecolor=black,border=0](0.4,0.5)(0.4,0.18)
  \psset{linewidth=0.9pt,linecolor=black,arrowscale=1.4,arrowinset=0.15}
  \psline{->}(0.2,0.37)(0.3,0.34)
  \rput[bl]{0}(0.0,0.0){$\omega_a$}
\endpspicture
\label{eq:T_loop}
,
\end{equation}
is equivalent to the application of $m$ twists to all the topological charge lines passing through the loop, and $\tilde{\Pr}_{AC}(a)$ is the probability of twisted charge measurement outcome $a$ by the twisted interferometer. The $\tau^m$-loops here correspond to the $\gamma$, $\bar{\gamma}$, $\beta$, and $\bar{\beta}$ curves in the handle bodies.

\subsection{Topological Understanding}
\label{sec:top_understanding}

%%%%%%%%%%%%%%%%%%%%%%%%%%%%%%%%%%%%%%%%%%%%%%%%%%%%%%%%%%%%%%%%%%%%%%%%%%%%%%%%%%%%%%%%%%%%%%
\begin{figure}[t!]
  \labellist
  % labels for (a)
  \pinlabel $\text{(a)}$ at 0 1450
  \pinlabel $C_1$ at 100 1180
  \pinlabel $A$ at 420 1180
  \pinlabel $C_2$ at 750 1180
  \pinlabel $t_1$ at 280 1180
  \pinlabel $t_2$ at 550 1180
  % labels for (b)
  \pinlabel $\text{(b)}$ at 1000 1450
  \pinlabel $C_1$ at 1050 1320
  \pinlabel $A$ at 1470 1500
  \pinlabel $C_2$ at 1900 1320
  \pinlabel $e_1$ at 1370 1100
  \pinlabel $e_2$ at 1570 1100
  \pinlabel $C_1^{\prime}$ at 1040 1030
  \pinlabel $A^{\prime}$ at 1470 860
  \pinlabel $C_2^{\prime}$ at 1900 1030
  \pinlabel $\text{top}$ at 1830 1460
  % labels for (c)
  \pinlabel $C_1$ at 10 530
  \pinlabel $A$ at 290 790
  \pinlabel $C_2$ at 570 530
  \pinlabel $C_1^{\prime}$ at 10 210
  \pinlabel $A^{\prime}$ at 290 -30
  \pinlabel $C_2^{\prime}$ at 570 210
  \pinlabel $e_1$ at 220 320
  \pinlabel $e_2$ at 410 320
  \pinlabel $\text{top}$ at 530 720
  \pinlabel $\text{top}$ at 500 0
  \pinlabel $\text{5 faces}$ at 550 100
  \pinlabel $\text{(c)}$ at 0 730
  % labels for (d)
  \pinlabel $C_1$ at 660 530
  \pinlabel $A$ at 940 790
  \pinlabel $C_2$ at 1220 530
  \pinlabel $C_1^{\prime}$ at 660 210
  \pinlabel $A^{\prime}$ at 940 -30
  \pinlabel $C_2^{\prime}$ at 1220 210
  \pinlabel $e_1$ at 870 320
  \pinlabel $e_2$ at 1060 320
  \pinlabel $\omega_0$ at 870 600
  \pinlabel $\omega_0$ at 1060 580
  \pinlabel $\text{(d)}$ at 650 730
  % labels for (e)
  \pinlabel $C_1$ at 1310 530
  \pinlabel $A$ at 1590 790
  \pinlabel $C_2$ at 1870 530
  \pinlabel $C_1^{\prime}$ at 1310 210
  \pinlabel $A^{\prime}$ at 1590 -30
  \pinlabel $C_2^{\prime}$ at 1870 210
  \pinlabel $e_1$ at 1520 320
  \pinlabel $e_2$ at 1710 320
  \pinlabel $\omega_0$ at 1520 600
  \pinlabel $\omega_0$ at 1710 580
  \pinlabel $\omega_a$ at 1635 570
  \pinlabel $\omega_a$ at 1550 170
  \pinlabel $\text{(e)}$ at 1300 730
  \endlabellist
  \begin{center}
  \includegraphics[width=0.9\textwidth]{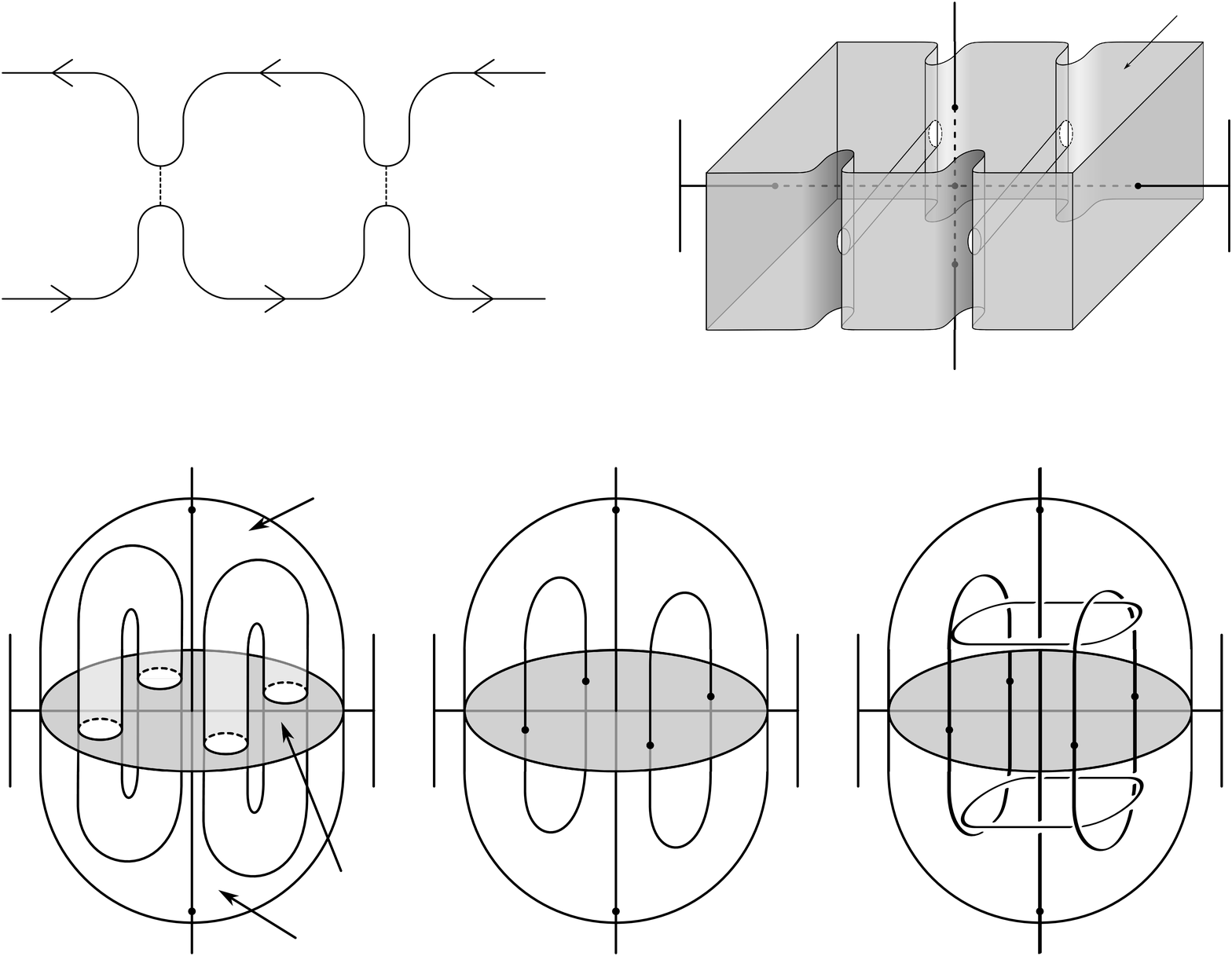}\vspace{0.75cm}
  \caption{(a) Spatial configuration of a fractional quantum Hall double-point contact interferometer. (b) Space-time description, including tunneling events, represented by evacuations of topological fluid. (c) Doubling the space-time along the five shaded faces (bottom and 4 sides) and gluing each tube to its mirror image. (d) Collapsing the tubes in (c) to Wilson lines, labeled by $\omega_0$. (e) Doubling space-time, gluing tubes to their cross components (i.e. the interference terms), and collapsing the tubes gives rise to the $\omega_a$-loops.}
  \label{fig:hall_bar}
  \end{center}
\end{figure}
%%%%%%%%%%%%%%%%%%%%%%%%%%%%%%%%%%%%%%%%%%%%%%%%%%%%%%%%%%%%%%%%%%%%%%%%%%%%%%%%%%%%%%%%%%%%%%%%%%%%%%%%%%%%%%%%%%%

We have used local diagrammatic calculations~\cite{Bonderson07b,Bonderson07c,Bonderson13b} as input to topological machinery. The output has been the surgical operation described in Section~\ref{sec:twist_consequence}. It is also possible, retrospectively, to give an illuminating, if not rigorous, topological explanation of the rules derived in~\cite{Bonderson13b} through the diagrammatic method. To give this explanation, it is convenient to think of a fractional quantum Hall double point-contact (Fabrey-P\'{e}rot) interferometry in the low tunneling limit (where its effect is essentially the same as the idealized Mach-Zehnder). In Fig.~\ref{fig:hall_bar}, we draw the space-time history of the topological fluid. We take the point of view that the fluid has been ``evacuated'' along tubes representing the collective tunneling path of the probes and that, because a large and indeterminate number of probes have passed, we know nothing about the effective topological charge on the meridians of these tubes. (The meridional topological charge could be any fusion product of multiple probe anyons. The probe quasiparticles in most cases will have small effective mass and correspond to edge theory tunneling operators with lowest scaling exponents (conformal dimensions), from which all other quasiparticles can be generated as composites.) To produce the manifold (with framed Wilson lines) $X$ corresponding to the partition function $Z$, we should double the space-time history along its boundary and past, and then further trace out unknown degrees of freedom on the meridians of the tubes by gluing each tube boundary to its mirror image. This last step folds each longitude loop $\gamma$ over itself to become an arc $\alpha$. Topologically, this is precisely what a zero-framed surgery accomplishes. The latter provides a disk $\Delta$ for each longitude loop $\gamma$ to bound, but after providing $\Delta$, it is topologically equivalent to then projecting $\Delta$ to one of its coordinates, resulting in the arc $\alpha$. The surgeries are encoded by the $\omega_0$-loops in Fig.~\ref{fig:hall_bar}(d) and (e). Gluing a tube to its mirror image means that each longitudinal circle $\gamma$ = (transversal arc) $\cup$ (mirror image transversal arc) gets collapsed to a single arc $\alpha$. Topologically, this is equivalent to providing a disk of space-time topological fluid to span across each longitudinal circle: $\partial B^2\times\ast, \ast\in\partial B^2_{\text{second factor}}$.

This explains, via the surgery/handle attachment picture, the passage from (c) to (d) in Fig.~\ref{fig:hall_bar}. We can represent the glued tubes as two new Wilson loops labeled by $\omega_0$, as explained in Section~\ref{sec:surgery_handle}. The final frame Fig.~\ref{fig:hall_bar}(e) includes the $\omega_a$-loops reflecting what the interferometer was ``intended'' to do, i.e. project $A$ into topological charge sector $a$ by measuring the interference term between the two tunneling paths. From this point of view, the $\omega_0$-loops are an ``unintended'' consequence of running the interferometer: tunneling the stream of probes anyons $B$ ``inadvertently'' decohered system $A$ from it complementary anyons $C_1$ and $C_2$.

%%%%%%%%%%%%%%%%%%%%%%%%%%%%%%%%
\subsection{The Ising Theory}
\label{sec:Ising_theory}
%%%%%%%%%%%%%%%%%%%%%%%%%%%%%%%%%

The twisted interferometry analysis represents a completely general tool for investigating the effects in general (2+1)D anyonic systems. However, we are primarily interested in the application for the Ising-type TQFTs, as these are the most physically practical non-Abelian anyonic systems to physically realize and are also the only examples we know (so far) that twisted interferometry provides an enhancement of computational utility. Ising TQFTs have topological charges $I$ (vacuum), $\sigma$ (non-Abelian anyon), and $\psi$ (fermion), where the $\sigma$ anyon should have a (statistical) twist factor $\theta_\sigma = e^{2\pi ix/16}$ for $x$ odd. This is the crucial $T$-matrix entry. In our calculation, we take $x=1$, but the other choices yield similarly useful results. The remainder of this paper is focused on this case.

For convenience, we recall the fusion and braiding properties of the Ising MTC
\begin{equation*}
\begin{tabular}{|l|l|}
\hline
\multicolumn{2}{|l|}{$\mathcal{C}=\left\{I,\sigma,\psi \right\}, \quad I\times a=a,\quad \sigma \times \sigma =I+\psi,\quad \sigma \times \psi=\sigma,\quad \psi \times \psi=I$}
\\ \hline
\multicolumn{2}{|l|}{$\left[ F_{\sigma}^{\sigma \sigma \sigma}\right] _{ef}=
\left[ F_{\sigma \sigma}^{\sigma \sigma}\right] _{ef}=
\left[
\begin{array}{rr}
\frac{1}{\sqrt{2}} & \frac{1}{\sqrt{2}} \\
\frac{1}{\sqrt{2}} & \frac{-1}{\sqrt{2}}%
\end{array}\right] _{ef}^{\phantom{T}}$} \\
\multicolumn{2}{|l|}{$\left[ F_{\psi}^{\sigma \psi \sigma}\right] _{\sigma \sigma}=%
\left[ F_{\sigma}^{\psi \sigma \psi}\right] _{\sigma \sigma_{\phantom{j}}}\!\!=
\left[ F_{\psi \sigma}^{\sigma \psi}\right] _{\sigma \sigma}=
\left[ F_{\sigma \psi}^{\psi \sigma}\right] _{\sigma \sigma}=-1 $} \\ \hline
\multicolumn{2}{|l|}{
$R_{I}^{\sigma \sigma}=e^{-i\frac{\pi }{8}},\quad R_{\psi}^{\sigma \sigma}=e^{i\frac{3\pi }{8}},
\quad R_{\sigma}^{\sigma \psi}=R_{\sigma}^{\psi \sigma}=e^{-i\frac{\pi }{2}},\quad R_{I}^{\psi \psi}=-1$} \\ \hline
$S=\frac{1}{2}\left[
\begin{array}{rrr}
1 & \sqrt{2} & 1 \\
\sqrt{2} & 0 & -\sqrt{2} \\
1 & -\sqrt{2} & 1%
\end{array}%
\right]^{\phantom{T}}_{\phantom{j}} $ & $M=\left[
\begin{array}{rrr}
1 & 1 & 1 \\
1 & 0 & -1 \\
1 & -1 & 1%
\end{array}%
\right] $ \\ \hline
$d_{I}=d_{\psi}=1,\quad d_{\sigma_{\phantom{j}}}\!\!=\sqrt{2}, \quad \mathcal{D}=2$ & $\theta _{I}=1,\quad \theta
_{\sigma}=e^{i\frac{\pi }{8}},\quad \theta _{\psi}=-1$ \\ \hline
\end{tabular}%
\end{equation*}%
The $F$-symbols and $R$-symbols not listed here are trivial, meaning they are equal to $1$ if allowed by the fusion rules.

%%%%%%%%%%%%%%%%%%%%%%%%%%%%%%%%%%%%%%%%%%%%%%%%%%%%%%%%%%%%%%%%%%%%%%%%%%%%%%%%%%%
\begin{figure}[t!]
  \labellist
  \pinlabel $\omega_a$ at 720 400
  \pinlabel $\omega_a$ at 620 30
  \pinlabel $\text{(a)}$ at 300 -60
  \pinlabel $\omega_a$ at 1550 400
  \pinlabel $\omega_a$ at 1450 30
  \pinlabel $\text{(b)}$ at 1120 -60
  \endlabellist
  \begin{center}
  \includegraphics[width=0.8\textwidth]{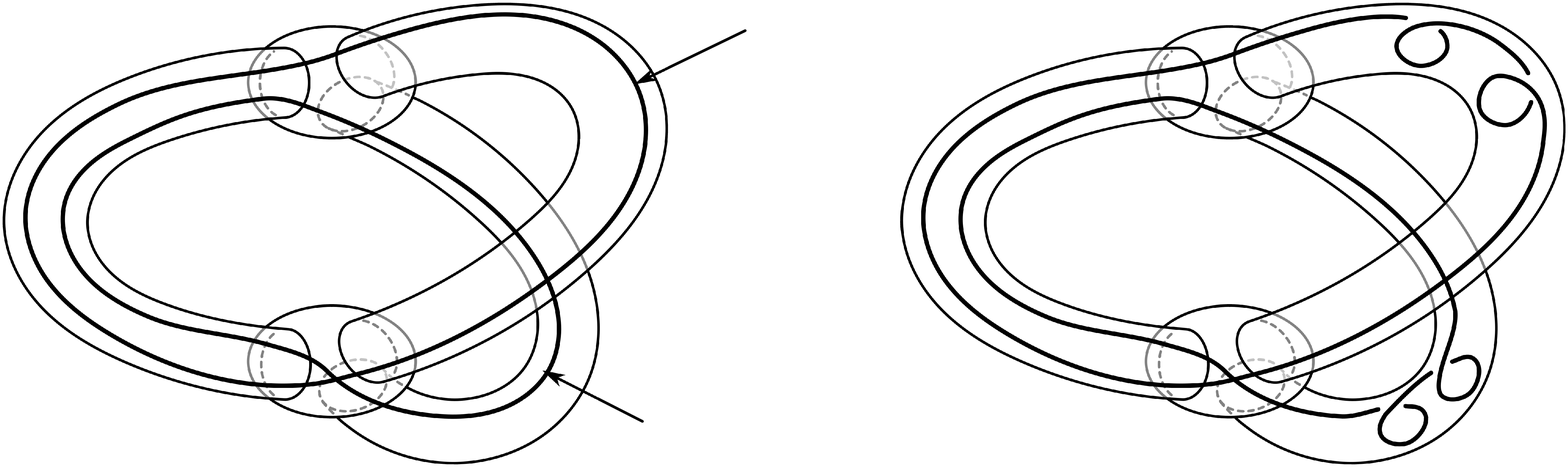}\vspace{0.5cm}
  \caption{Genus $2$ handle bodies for (a) untwisted and (b) twisted interferometry using Ising anyons when $a = I$ or $\psi$.}
  \label{fig:a_eq}
  \end{center}
\end{figure}
%%%%%%%%%%%%%%%%%%%%%%%%%%%%%%%%%%%%%%%%%%%%%%%%%%%%%%%%%%%%%%%%%%%%%%%%%%%%%%%%%%%%%%

The identity in Eq.~(\ref{id:loop_slide}) simplifies Figs.~\ref{fig:handle_bodies}(a) and \ref{fig:repositioned}(a) in the cases where we have a priori information (as will be present in the qubit context) that the topological charge $A$ (corresponding to the fusion channel of a pair of anyons from a 4 anyon topological qubit) is a linear combination of $I$ and $\psi$, so that the only possible $\omega_a$-labeled Wilson loops will have $a=I$ or $\psi$ (here we write $I$ for $0$).
This is exhibited in Fig.~\ref{fig:a_eq}, where we show the corresponding simplifications of Figs.~\ref{fig:handle_bodies}(a) and \ref{fig:repositioned}(a) for the Ising theory. In particular, the $\omega_0$-loop in those figures is redundant. This can be seen from the following argument. Using the fact that $\omega_a$ is idempotent, the upper $\omega_a$-loop can be replaced with two parallel $\omega_a$-loops, without changing the partition function $Z$. Next take one of the newly created upper $\omega_a$-loops and slide it over the lower $\omega_a$-loop using Eq.~(\ref{id:loop_slide}). The resulting loop, which now is labeled with an $\omega_0$, may finally be isotopied into the position of the $\omega_0$-loop in Figs.~\ref{fig:handle_bodies}(a) and \ref{fig:repositioned}(a). Thus, these configurations of $\omega$-loops are equivalent, demonstrating the redundance of the $\omega_0$-loop. An analogous argument similarly shows that only the $\omega_a$-loops need be considered for Ising anyons with $a=I$ or $\psi$ in Figs.~\ref{fig:handle_bodies}(b) and \ref{fig:repositioned}(b), when the complementary anyons are in two regions $C_1$ and $C_2$.

The conclusion is that when measuring topological qubits in Ising-like theories, it is harmless to omit the surgery (i.e. the $\omega_0$-loop) representing decoherence from a connected environment $C$. In the untwisted case, interferometry only gives projective measurement of the topological charge, with no decoherence of anyonic entanglement. In other words, the interferometry measurement superoperator takes pure states to pure states. This simplifies the calculation, allowing us to work with a single, rather than a doubled copy of space-time, since no surgery loops traverse the two factors.

In the case of two twists, $(r,l) = (2,0)$, as we will compute in Section~\ref{sec:double_twist}, the twisted interferometer (using probes with $b=\sigma$) acts on a state $\ket{\Psi} = \alpha\ket{I} + \beta\ket{\psi}$ by sending it to $\ket{\Psi'} = (1+e^{-2\pi i/8})\alpha\ket{I} + (1-e^{2\pi i/8})\beta\ket{\psi}$, if the ``twisted measurement outcome'' is charge $a=I$ and to $\ket{\Psi'} = (1-e^{2\pi i/8})\alpha\ket{I} + (1+e^{-2\pi i/8})\beta\ket{\psi}$ for $a=\psi$. Similarly, on the level of density matrices, for the initial target system density matrix~\footnote{The expression in terms of the qubit density matrix $\rho$ use the qubit basis states given by $\left| 0 \right\rangle = \left| I,I;I \right\rangle$ and $\left| 1 \right\rangle = \left| \psi,\psi;I \right\rangle$.}
\begin{equation}
\rho^{AC} = \sum\limits_{a,a^{\prime } = I,\psi } \rho _{\left( a,a;I \right) \left( a^{\prime },a^{\prime };I\right)}^{AC}
 \left| a,a;I \right\rangle \left\langle a^{\prime },a^{\prime };I \right| = \left[
\begin{array}{rr}
\rho _{00}  &  \rho _{01} \\
\rho _{10}  &  \rho _{11 }
\end{array}
\right]
,
\label{eq:Ising_initial_state}
\end{equation}
the outcome after twisted interferometry with outcomes $a=I$ or $\psi$ are, respectively
and resulting (fixed state) density matrices
\begin{equation}
\tilde{\rho}_{I} = \frac{1}{\tilde{\Pr}_{AC} \left( I \right)}
\left[
\begin{array}{cc}
\cos^2 \left( \frac{\pi}{8} \right) \rho _{00}  &  i \cos \left(  \frac{ \pi}{8} \right) \sin \left(  \frac{ \pi}{8} \right) \rho _{01} \\
-i \cos \left(  \frac{ \pi}{8} \right) \sin \left(  \frac{ \pi}{8} \right) \rho _{10}  & \sin^2 \left(  \frac{ \pi}{8} \right)   \rho _{11 }
\end{array}
\right]
,
\label{eq:rho_I}
\end{equation}
\begin{equation}
\tilde{\rho}_{\psi} = \frac{1}{\tilde{\Pr}_{AC} \left( \psi \right)}
\left[
\begin{array}{cc}
\sin^2 \left(  \frac{ \pi}{8} \right) \rho _{00}  &  -i \cos \left(  \frac{ \pi}{8} \right) \sin \left(  \frac{ \pi}{8} \right) \rho _{01} \\
i \cos \left(  \frac{ \pi}{8} \right) \sin \left(  \frac{ \pi}{8} \right) \rho _{10}  & \cos^2 \left(  \frac{\pi}{8} \right)   \rho _{11 }
\end{array}
\right]
,
\label{eq:rho_psi}
\end{equation}
with corresponding probabilities
\begin{eqnarray}
\tilde{\Pr}_{AC} \left( I \right) &=& \cos^2 \left( \pi/8 \right) \rho_{00} +  \sin^2 \left( \pi/8 \right) \rho_{11},
\label{eq:prob_I} \\
\tilde{\Pr}_{AC} \left( \psi \right) &=& \sin^2 \left( \pi/8 \right) \rho_{00} +  \cos^2 \left( \pi/8 \right) \rho_{11}
\label{eq:prob_psi}
\end{eqnarray}

Importantly, in the twisted case, regardless of whether the redundant $\omega_0$-loop is included in the diagram, there is no decoherence of anyonic entanglement between the target anyons $A$ and their complementary anyons $C$, and the final state may possess coherent superposition of topological charges and anyonic entanglement between $A$ and $C$. This seemingly paradoxical fact is explained by the fact that the $\omega_0$-loop, which normally causes decoherence between $A$ and $C$ for the untwisted case, is (double) twisted around the two $\omega_a$-loops. When twisted in this manner, the $\omega_0$-loop no longer separates the target system $A$ from $C$.

%%%%%%%%%%%%%%%%%%%%%%%%%%%%%%%%%%%%%%%%%%%%%%%%%%%%%%%%
\section{The Double Twisted Interferometer in Ising Systems}
\label{sec:double_twist}
%%%%%%%%%%%%%%%%%%%%%%%%%%%%%%%%%%%%%%%%%%%%%%%%%%%%%%%%

In this section, we calculate the asymptotic effect of running a twisted interferometer with two twists in one arm, as indicated in Fig.~\ref{fig:Double_Twisted_int}, for a system with Ising non-Abelian anyons. We are interested in a configuration where the anyons $A$ are composed of a pair of $\sigma$ anyons, which may be part of a topological qubit (requiring at least two complementary $\sigma$ anyons in $C$) and can have collective fusion channel values $I$ and $\psi$. The probe quasiparticles are assumed to carry topological charge $b=\sigma$. With appropriate assumptions, the analysis also extends to other twisted interferometer designs, such as those described in~\cite{Bonderson13b}. In the first two subsections, we review general TQFT technology. The effect of the twisting is computed in the final subsection.

\subsection{Gluing 3-Manifolds and Tensor Contractions}
\label{sec:tensor_contractions}

The basic structure of a TQFT is a functor that assigns Hilbert spaces $H(\Sigma)$ to a surface $\Sigma$ and partition functions $Z(M)$ to 3-manifolds $M$. If $M$ is closed (compact and without boundary, $\partial M = \varnothing$), then the partition function $Z(M)$ is a scalar. If $M$ has a single boundary component $\Sigma$, then $Z(M)\in H(\Sigma)$. If $\partial M$ is divided into two pieces, say incoming and outgoing with respect to the orientation of $M$, then $Z(M)\in\text{Hom}(H(\Sigma_{\text{in}}),H(\Sigma_{\text{out}}))$. The division of $\partial M$ into pieces may be according to components, but this is not essential. Several boundary components may be grouped into one piece and one component may be cut apart along non-intersecting simple closed curves (SCCs) into two or more pieces. When SCCs are present, the boundary pieces $\Sigma_i$ themselves have boundary and the appropriate Hilbert space $H(\Sigma_i)$ is a direct sum (scaled according to quantum dimensions) of all admissible topological charge labelings of the boundary components. In any case, if $\partial M$ is divided into $k$ pieces, the TQFT assigns a $k$-tensor to $M$. Orientation conventions determine which indices are covariant and which are contravariant.

The ``Atiyah axiom,'' which is the fundamental gluing relation, is:
\begin{equation}
\label{eq:atgah_axiom}
Z(M\cup N) = \langle Z(M),Z(N)\rangle
\end{equation}
where $M$ and $N$ are glued over a common piece of boundary and the symbol $\langle\cdot,\cdot\rangle$, suggestive of the inner product, means contract the tensors along the index associated to the glued piece of boundary.

\begin{figure}[t!]
  \labellist \small\hair 2pt
  \pinlabel $i$ at -10 180
  \pinlabel $j$ at -10 10
  \pinlabel $k$ at 150 110
  \pinlabel $k$ at 215 110
  \pinlabel $i$ at 500 180
  \pinlabel $j$ at 500 10
  \pinlabel $\mathcal{T}_{ijk}$ at 100 10
  \pinlabel $\textrm{3-tensor}$ at 100 -10
  \pinlabel $v_{k}$ at 280 10
  \pinlabel $\textrm{co-vector}$ at 280 -10
  \pinlabel $\sum_{k}\mathcal{T}_{ijk}v_{k}$ at 600 10
  \pinlabel $\textrm{2-tensor}$ at 600 -15
  \pinlabel $\textrm{contraction}$ at 420 100
  \pinlabel \Huge{$\leadsto$} at 420 70
  \endlabellist
  \begin{center}
  \includegraphics[width=4in]{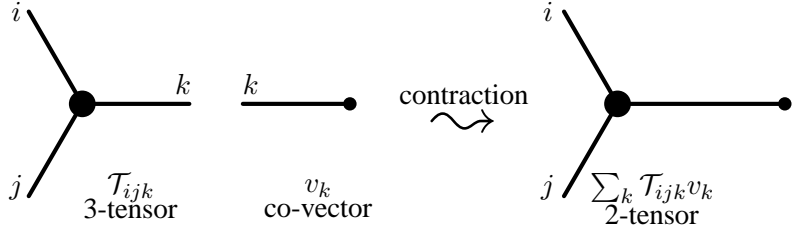}\vspace{0.5cm}
  \caption{Tensor contraction.}
  \label{fig:tensor_construction}
  \end{center}
\end{figure}

For example, if $M$ has among its boundary components a torus $T$ and $N=S^1\times D^2$ is a solid torus with boundary identified to $T$ ($N$ may contain a charged Wilson loop at its core), then $Z(M\cup N)$ is obtained as a tensor contraction as in Fig~\ref{fig:tensor_construction}.

%%%%%%%%%%%%%%%%%%%%%%%%%%%%%%%%%%%%%%%%%%%%%%%%%%%%%%%%%%%%%%%%%%%%%%%%%
%%%%%%%%%%%%%%%%%%%%%%%%%%%%%%%%%%%%%%%%%%%%%%%%%%%%%%%%%%%%%%%%%%%%%%%%%%%%

As we run interferometers (twisted or untwisted), we are effectively measuring topological charge along a longitudinal loop $\gamma\subset T$ in a torus boundary component of a topological space-time fluid. A cavity $N$ bounded by $T$ arises as the stream of probes $B$ annihilates the topological fluid along the interferometry loop, as discussed in Section~\ref{sec:top_understanding}. The measurement outcome effectively replaces the deleted solid torus and boundary $(N,T)$ by a new one $(N^{\prime},T^{\prime})$ with meridian ($T^{\prime}$) glued to longitude ($T$). The new solid torus $N^{\prime}$, within this effective description, enforces the measured charge $a$. To do this it contains a Wilson loop labeled by topological charge $a$ at its core.

Given a TQFT, one should think of a given 3-manifold $M$ with boundary as a family of tensors that depend on how its boundary is divided into pieces. In the next section, we see that this is already a rich discussion when the TQFT is the Ising theory, $M$ is a solid torus and $\partial M$ is divided into two annuli, but \emph{partitioned} in a variety of ways. For the Ising TQFT (with connected complement $C$), the effect of interferometric measurement is merely a Dehn surgery (with $\omega_a$ Wilson loops having $a=I$ or $\psi$, depending on measurement outcome) effecting a tensor contraction with the observed state.

\subsection{TQFTs: A Fixed 3-Manifold Yields Many Tensors According to its Boundary Decomposition}
\label{sec:many_tensors}

\begin{figure}[t!]
  \labellist
  \pinlabel $\text{(a)}$ at -15 110
  \pinlabel $\text{(b)}$ at 250 110
  \pinlabel $\text{(c)}$ at 515 110
  \endlabellist
  \begin{center}
  \includegraphics[width=4in]{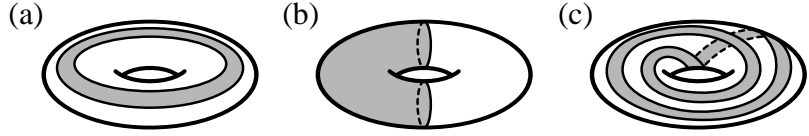}\vspace{0.5cm}
  \caption{Three different decompositions of the 2D torus boundary of a 3D solid torus. In each of these examples, the boundary torus is partitioned into two annuli, which are colored white and grey, respectively.}
  \label{fig:torus_example}
  \end{center}
\end{figure}

The 3-manifold $M$ plays the role of the tensor $\mathcal{T}$, but its valence is unspecified until the (2-manifold) boundary of $M$ is dissected into pieces. These pieces may be closed or themselves have a 1-manifold boundary, which specifies the index set for the tensor. The axioms for TQFTs strongly restrict which tensors arise as the boundary decomposition of $M$ is varied. For a key example, take $M$ to be a solid torus $S^1\times D^2$  and the Ising TQFT (see Section~\ref{sec:Ising_theory} for a summary of the Ising TQFT rules). Decomposing the 2D torus boundary $\partial M$ into annuli $A$ and $B$ ($\partial M = A\cup B$) as shown in Fig.~\ref{fig:torus_example} yields three different matrices (2-tensors), with indices corresponding to the $I$, $\sigma$, and $\psi$ topological charge basis along the two loops (1-manifolds) of $A\cap B$. These boundary partitions will be useful, so we sketch how the calculations are done for the examples in Fig.~\ref{fig:torus_example}.

%%%%%%%%%%%%%%%%%%%%%%%%%%%%%%%%%%%%%%%%%%%%%%%%%%%%%%%%%%%%%%%%%%%%%%%%%%%
%%%%%%%%%%%%%%%%%%%%%%%%%%%%%%%%%%%%%%%%%%%%%%%%%%%%%%%%%%%%%%%%%%%%%%%%%%%%%

For the boundary partition in Fig.~\ref{fig:torus_example}(a), the result is axiomatic: products correspond to identity morphisms. The identity operator ``glues up'' to become the vector (1-index tensor)
\begin{equation}
v_l = \begin{pmatrix}
\frac{d_I}{\mathcal{D}} \\ \frac{d_{\sigma}}{\mathcal{D}} \\ \frac{d_{\psi}}{\mathcal{D}}
\end{pmatrix}
=\begin{pmatrix}
\frac{1}{2} \\ \frac{\sqrt{2}}{2} \\ \frac{1}{2}
\end{pmatrix}
\end{equation}
in the vector space $V_l(T)$ corresponding to the longitudinal basis. The corresponding operator $\mathcal{O}_{l} = \mathbb{I}$ is obtained by placing the entries of the vector on the diagonal of the matrix and dividing by $S_{I,a} = \frac{d_a}{\mathcal{D}}$ to obtain the proper normalization, i.e.
\begin{equation}
[ \mathcal{O}]_{a,b} = \frac{ [v]_a }{S_{I,a}} \delta_{a,b}
.
\end{equation}

The result for the boundary partition in Fig.~\ref{fig:torus_example}(b) can be obtained from (a) by applying the modular $S$-transformation
\begin{equation}
S=\frac{1}{2}\left[
\begin{array}{ccc}
1 & \sqrt{2} & 1 \\
\sqrt{2} & 0 & -\sqrt{2} \\
1 & -\sqrt{2} & 1
\end{array}
\right]
,
\end{equation}
which transforms between the longitudinal and meridional bases. In this way, we obtain
\begin{equation}
v_m = S (v_l) =
\begin{pmatrix}
1 \\ 0 \\ 0
\end{pmatrix}
.
\end{equation}
The corresponding operator is
\begin{equation}
\mathcal{O}_{m} = \left[
\begin{array}{ccc}
2 & 0 & 0\\
0 & 0 & 0\\
0 & 0 & 0
\end{array}
\right]
.
\end{equation}

Finally, to compute the result for Fig.~\ref{fig:torus_example}(c), we note that
\begin{equation}
B = S T^2 S^{-1} = \left[
\begin{array}{ccc}
\frac{1+\omega}{2} & 0 & \frac{1-\omega}{2}\\
0 & 1 & 0\\
\frac{1-\omega}{2} & 0 & \frac{1+\omega}{2}
\end{array}
\right]
,
\end{equation}
with $\omega = e^{i 2\pi /8}$, is the modular transformation sending (b) to (c), where
\begin{equation}
T=\left[
\begin{array}{ccc}
1 & 0 & 0 \\
0 & e^{i \frac{2 \pi}{16}} & 0 \\
0 & 0 & -1
\end{array}
\right]
\end{equation}
is the modular Dehn twist transformation, which cuts open the torus along the meridian and glues it back together with a $2 \pi$ twist. Then, in this twisted basis ($t$), the vector for Fig.~\ref{fig:torus_example}(c) is
\begin{equation}
v_t = B(v_m) =
\begin{pmatrix}
\frac{1+\omega}{2}\\0\\\frac{1-\omega}{2}
\end{pmatrix}
.
\end{equation}
The corresponding operator is
\begin{equation}
\mathcal{O}_{t} = \left[
\begin{array}{ccc}
1+\omega & 0 & 0\\
0 & 0 & 0\\
0 & 0 & 1-\omega
\end{array}
\right]
,
\end{equation}
where, as mentioned, we divided entries by $S_{I,a} = \frac{d_a}{\mathcal{D}}$ to obtain the proper normalization.

We record also the vector and operator associated with a case (c$^{\prime}$), which is the same boundary data as case (c), but with the solid torus containing a $\psi$-charge Wilson loop running along its core. In case (c$^{\prime}$), we should now apply the above to the vector $v_m^{\prime} = (0,0,1)^{T}$ corresponding to meridinal charge $\psi$. This gives
\begin{equation}
v_t = B (v_m^{\prime}) = \begin{pmatrix}
\frac{1-\omega}{2}\\0\\\frac{1+\omega}{2}
\end{pmatrix}
.
\end{equation}

Thus, the corresponding operator is
\begin{equation}
\label{eq:O_t}
\mathcal{O}_{t} = \left[
\begin{array}{ccc}
1-\omega & 0 & 0\\
0 & 0 & 0\\
0 & 0 & 1+\omega
\end{array}
\right]
.
\end{equation}

Gluing a 3 dimensional solid torus $D^2\times S^1 = D^2\times \partial D^2$ is the TQFT equivalent of tracing (summing over a repeated index). In our application, $D^2\times \partial D^2$ is a solid torus of space-time topological fluid glued into the cavity created by removing a solid torus ($D^2 \times \partial D^2$) neighborhood of the interferometry loop $\gamma$. The gluing should respect the framing on $\gamma$.

%%%%%%%%%%%%%%%%%%%%%%%%%%%%%%%%%%%%%%%%%%%%%%%%%%%%%%%%%%%%%%%%%%%%%%
\begin{figure}[t!]
  \labellist
  \pinlabel $a$ at 560 180
  \pinlabel $D^2\times\partial D^2$ at 700 30
  \endlabellist
  \begin{center}
  \includegraphics[width=0.3\textwidth]{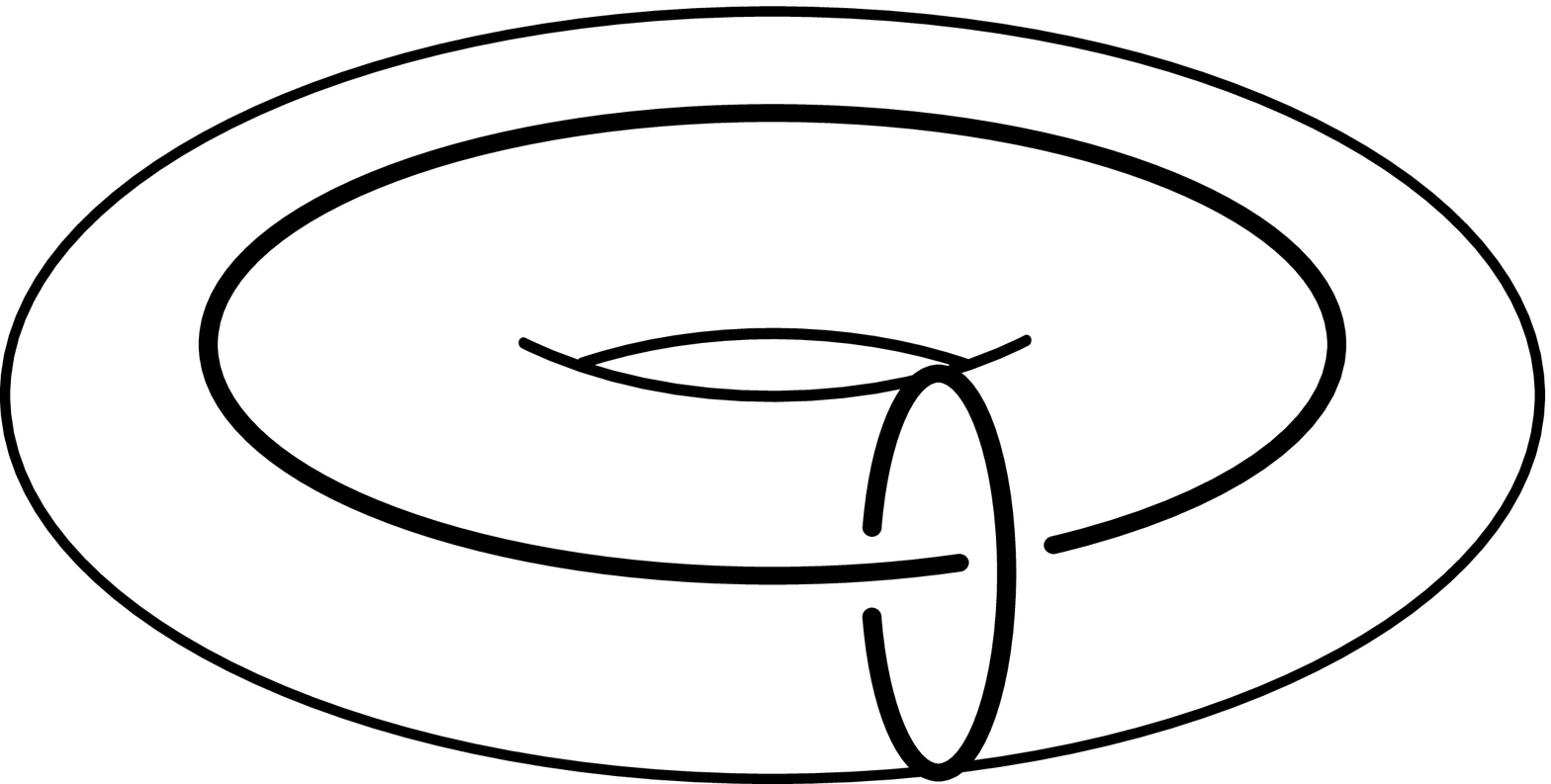}\vspace{0.5cm}
  \caption{Glue along the torus $T^2 = \partial D^2\times \partial D^2$, respecting the framing of $\gamma$. Here, $a = I$ or $\psi$, the measured charge on curve $\gamma$.}
  \label{fig:torus_gluing}
  \end{center}
\end{figure}
%%%%%%%%%%%%%%%%%%%%%%%%%%%%%%%%%%%%%%%%%%%%%%%%%%%%%%%%%%%%%%%%%%%%%%

%%%%%%%%%%%%%%%%%%%%%%%%%%%%%%%%%%%%%%%%%%%%%%%%%%%%%%%%%%%%%%%%%%%%%%%%
\begin{figure}[t!]
  \labellist \small\hair 2pt
  \pinlabel $M^3$ at 200 135
  \pinlabel $T$ at 355 190
  \pinlabel \Huge{$\leadsto$} at 500 125
  \pinlabel $\textrm{solid torus}$ at 1000 70
  \pinlabel $a$ at 900 190
  \pinlabel $\text{(a)}$ at 200 -70
  \pinlabel $\text{(b)}$ at 750 -70
  \endlabellist
  \begin{center}
  \includegraphics[width=3.5in,height=1.2in]{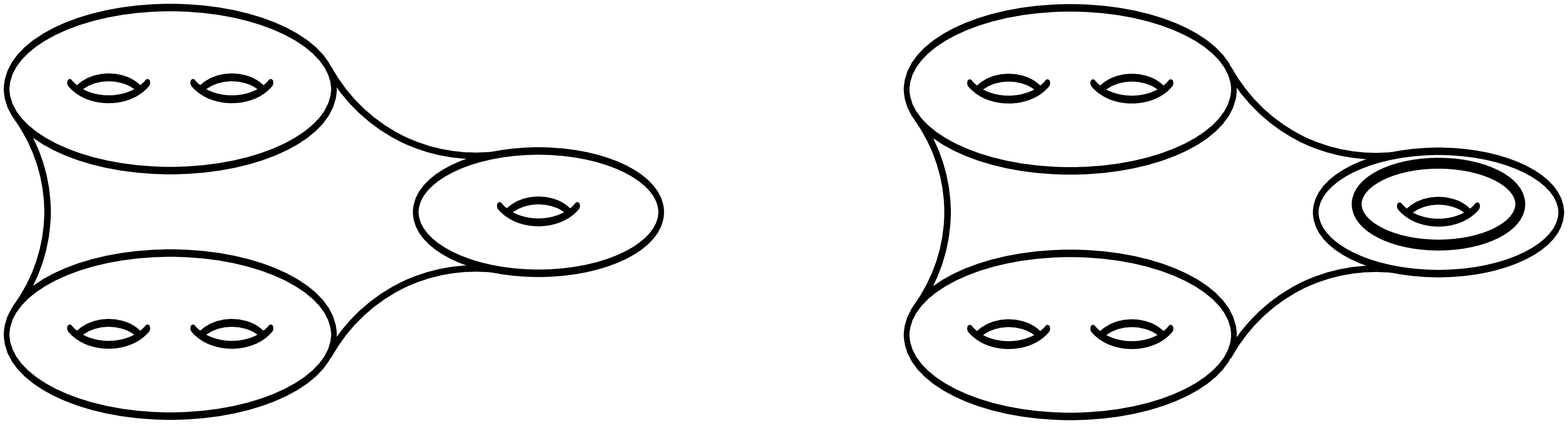}\vspace{0.75cm}
  \caption{Measurement with outcome observing topological charge $a$ along the curve $\gamma$. Note the analogy to Fig.~\ref{fig:tensor_construction}.}
  \label{fig:measuring_alpha}
  \end{center}
\end{figure}
%%%%%%%%%%%%%%%%%%%%%%%%%%%%%%%%%%%%%%%%%%%%%%%%%%%%%%%%%%%%%%%%%%%%%%%%%%%

The topological charge $a$ line at the core of the replacement solid torus is precisely the measurement outcome $a=I$ or $\psi$. (If the measured topological charge value is trivial $I$, the solid torus has no Wilson line.) Up to an overall scalar, which has no physical significance, measuring charge $a$ along the curve $\gamma$ is equivalent to deleting a $D^2 \times \partial D^2$ neighborhood of $\gamma$ and re-gluing $D^2 \times \partial D^2$ with $\ast\times\partial D^2$ matching the first normal frame vector to $\gamma$, $\ast\in\partial D^2$, and $0\times\partial D^2$ being a Wilson loop of charge $a$.
Thus a measurement of $a=I$ or $\psi$ Dehn fills a new solid torus near $\gamma$ with a Wilson loop of charge $a$ at its core.

%%%%%%%%%%%%%%%%%%%%%%%%%%%%%%%%%%%%%%%%%%%%%%%%%%%%%%
\subsection{Effect of Twisting: The $\pi/8$-phase gate}
\label{sec:effect_derivation}
%%%%%%%%%%%%%%%%%%%%%%%%%%%%%%%%%%%%%%%%%%%%%%%%%%%%%%%

In Section~\ref{sec:many_tensors}, we calculated the operator $\mathcal{O}_t$ associated to a solid torus with $(1,-2)$-twisted boundary, as shown in Fig.~\ref{fig:torus_example}(c), containing an $I$ or $\psi$ Wilson loop.
In the longitudinal basis, restricted to topological charge values $I$ and $\psi$, this was given by
\begin{equation}
\label{eq:qubit_operation}
\mathcal{O}_t = \left[
  \begin{matrix}
    1+\omega & 0\\
    0 & 1-\omega
  \end{matrix}
  \right]
  \quad\text{or}\quad
  \left[
  \begin{matrix}
    1-\omega & 0\\
    0 & 1+\omega
  \end{matrix}
  \right].
\end{equation}
according to whether the Wilson loop has charge $I$ or $\psi$.
This operator, together with Ising anyon braiding transformations and standard (untwisted) interferometry measurements, allows one to generate $\pi /8$-phase gates
\begin{equation}
R_{\frac{\pi}{4}} = \left[
\begin{matrix}
1 &0\\
0 & e^{i\pi/4}
\end{matrix}
\right]
.
\end{equation}
In particular, applying $\mathcal{O}_t$ to the state $\frac{1}{\sqrt{2}} \left( \ket{0} + \ket{1} \right) = H \ket{0} $, where the Hadamard operator
\begin{equation}
H = \frac{1}{\sqrt{2}} \left[
\begin{matrix}
1 & 1 \\
1 & -1
\end{matrix}
\right]
\end{equation}
can be obtained as a braiding transformation, generates the ``magic state''
\begin{equation}
\left| \mathcal{B}_{-\frac{\pi}{4} } \right\rangle =  H R_{\frac{\pi}{4}} H \left| 0 \right\rangle= \cos(\pi/8) \left| 0 \right\rangle - i \sin(\pi/8) \left| 1 \right\rangle
\end{equation}
(up to an overall scalar that is removed by normalization) or
\begin{equation}
\left| \mathcal{B}_{\frac{3\pi}{4} } \right\rangle =  \sigma^x H R_{\frac{\pi}{4}} H \left| 0 \right\rangle= \sin(\pi/8) \left| 0 \right\rangle + i \cos(\pi/8) \left| 1 \right\rangle
,
\end{equation}
depending on whether one uses the $I$ or $\psi$ operator $\mathcal{O}_t$.
Using Ising braiding gates and measurements, any magic state (such as these) can be transformed into $\pi /8$-phase gates.

In the untwisted context, the measurement imposes one of the two projections, in the basis of topological charge $I$ or $\psi$ enclosed in the untwisted interferometry loop, given by
\begin{equation}
\Pi_0 = \left[
\begin{matrix}
1&0\\
0&0
\end{matrix}
\right]
,
\end{equation}
if charge $I$ is observed and
\begin{equation}
\Pi_1 = \left[
\begin{matrix}
0&0\\
0&1
\end{matrix}
\right]
,
\end{equation}
if charge $\psi$ is observed. One might na\"ively expect the twisted interferometer to generate conjugates of $\Pi_0$ and $\Pi_1$, however, this is not correct because the matrices obtained have rank 2. Since no charge lines enter or leave the twisted interferometer (and we always assume there are no mobile charges) the twisted interferometry operator $\mathcal{O}_t$ must be diagonal in the $I,\psi$ basis of topological charge [which is a consistency check on Eq.~(\ref{eq:qubit_operation})].

%%%%%%%%%%%%%%%%%%%%%%%%%%%%%%%%%%%%%%%%%%%%%%%%%%%%%%%%%%%%%%%%%%%%%%%%%%%%%%%%%%
\begin{figure}[t!]
  \labellist \small\hair 2pt
  \pinlabel $l^{\prime}$ at 200 80
  \pinlabel \huge{$\approx$} at 350 130
  \pinlabel $l^{\prime}$ at 670 180
  \endlabellist
  \begin{center}
  \includegraphics[width=0.7\textwidth]{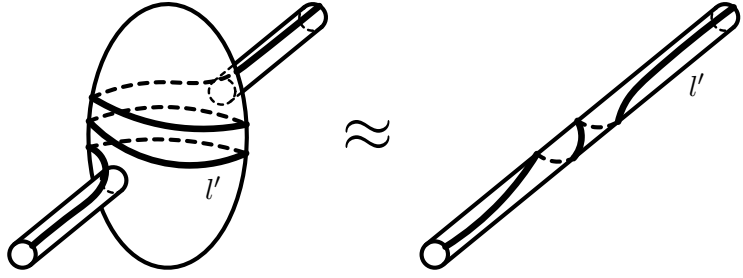}
  \caption{Twisted interferometry loop $l^{\prime}$ near vacuum island.}
  \label{fig:twisted_interferometry}
  \end{center}
\end{figure}
%%%%%%%%%%%%%%%%%%%%%%%%%%%%%%%%%%%%%%%%%%%%%%%%%%%%%%%%%%%%%%%%%%%%%%%%%%%%%%%%

\begin{figure}[t!]
  \labellist \small\hair 2pt
  \pinlabel $A$ at 70 260
  \pinlabel $l$ at 180 260
  \pinlabel $\gamma$ at 210 125
  \pinlabel $B$ at 70 50
  \pinlabel $=$ at 370 150
  \pinlabel $A$ at 445 155
  \pinlabel $B$ at 610 170
  \endlabellist
  \begin{center}
  \includegraphics[width=4.5in,height=2.5in]{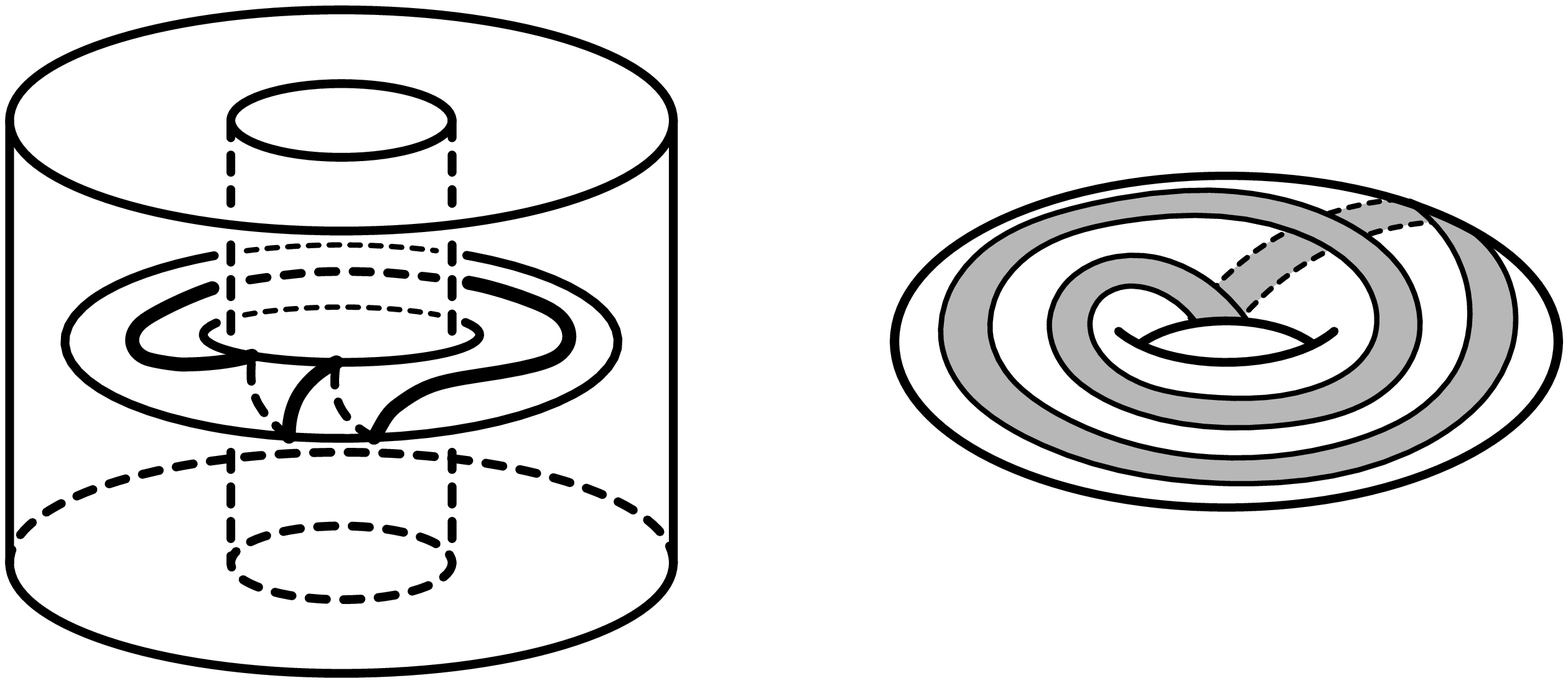}
  \caption{A change of coordinates using Dehn twists equates the solid torus describing the region of space-time in which the twisted interferometry takes place (left) with the solid torus with the previously analyzed boundary partition (right). The anyons being measured are contained within the missing core, i.e. are enclosed in the spatial plane (a time slice) by the loop $l$. The top annular boundary region $A$ on the left maps to the shaded twisted band $A$ on the surface of the solid torus on the right; the bottom annular boundary region $B$ on the left maps to the white twisted band $B$ on the surface of the solid torus on the right; the inner and outer vertical boundaries of the solid on the left map to the two boundaries (black lines) separating the regions $A$ and $B$ on the surface of the solid torus on the right. Measuring $l^{\prime}$ Dehn fills a solid torus on the right, so that $l^{\prime}$ bounds a disk or disk with $\psi$ anyon. If the measurement outcome is $a =I$ or $\psi$, there is a Wilson loop of charge $a$ at the core of the solid torus.}
  \label{fig:coordinate_change}
  \end{center}
\end{figure}

The relation between the twisted interferometric path and the boundary conditions of Fig.~\ref{fig:torus_example}(c) is show in Figs.~\ref{fig:twisted_interferometry} and \ref{fig:coordinate_change}.
In Fig.~\ref{fig:twisted_interferometry}, the two extra trips around the island or along the twisted track mean that measurement is applied along a topologically twisted $(1,-2)$ loop, which is related to the spatial perimeter of the interferometer $l$ by a change of coordinates described by $B = S^{-1}T^2S$. Referring to Figs.~\ref{fig:twisted_interferometry} and \ref{fig:coordinate_change}, we see that the two changes of coordinates described in Section~\ref{sec:many_tensors} computes $\mathcal{O}_t$, in the case of the two measurement outcomes $I$ or $\psi$.

Suppose $l$ is the outer boundary of a standard Ising qubit encoded in the $I$ and $\psi$ fusion channels of $\sigma$ anyons. Running a generically tuned doubly twisted interferometer (with $\sigma$ probe quasiparticles that are assumed to have negligible probe-probe interaction) around $l$ (equivalent to $\gamma$ in Fig.~\ref{fig:coordinate_change} via Fig.~\ref{fig:twisted_interferometry}) asymptotically realizes the $\mathcal{O}_t$ operator (up to exponentially suppressed corrections), which can be used to implement a $\pi/8$-phase gate.

%%%%%%%%%%%%%%%%%%%%%%%%%%%%%%%%%%%%%%%%%%%%%%%%%%%%%%%%%%%%%%%%%%%%%%%%%%%%%%%%%%
%%%%%%%%%%%%%%%%%%%%%%%%%%%%%%%%%%%%%%%%%%%%%%%%%%%%%%%%%%%%%%%%%%%%%%%%%%%%%%%%%%%%

%%%%%%%%%%%%%%%%%%%%%%%%%%%%%%%%%%%%%%%%%%%%%%%%%%%%%%%%
\section{Protocol for Direct Implementation of $\pi/8$-Phase Gate}
\label{sec:direct_gate}
%%%%%%%%%%%%%%%%%%%%%%%%%%%%%%%%%%%%%%%%%%%%%%%%%%%%%%%%

\begin{figure}[t!]
  \labellist \small\hair 2pt
  \pinlabel $\ket{\Psi}$ at 60 -10
  \pinlabel $t=0$ at 780 60
  \pinlabel $\sigma$ at 220 30
  \pinlabel $\sigma$ at 570 30
  \pinlabel $\sigma$ at 200 520
  \pinlabel $\sigma$ at 315 520
  \pinlabel $\alpha$ at 520 520
  \pinlabel $\gamma$ at 75 280
  \pinlabel $\ket{\Psi^{\prime}}$ at 50 580
  \pinlabel $t=1$ at 780 520
  \pinlabel \rotatebox{90}{vacuum} at 560 300
  \pinlabel vacuum at 280 165
  \pinlabel \rotatebox{90}{$\underrightarrow{\quad\text{time}\quad}$} at 770 300
  \endlabellist
  \begin{center}
  \includegraphics[width=4in]{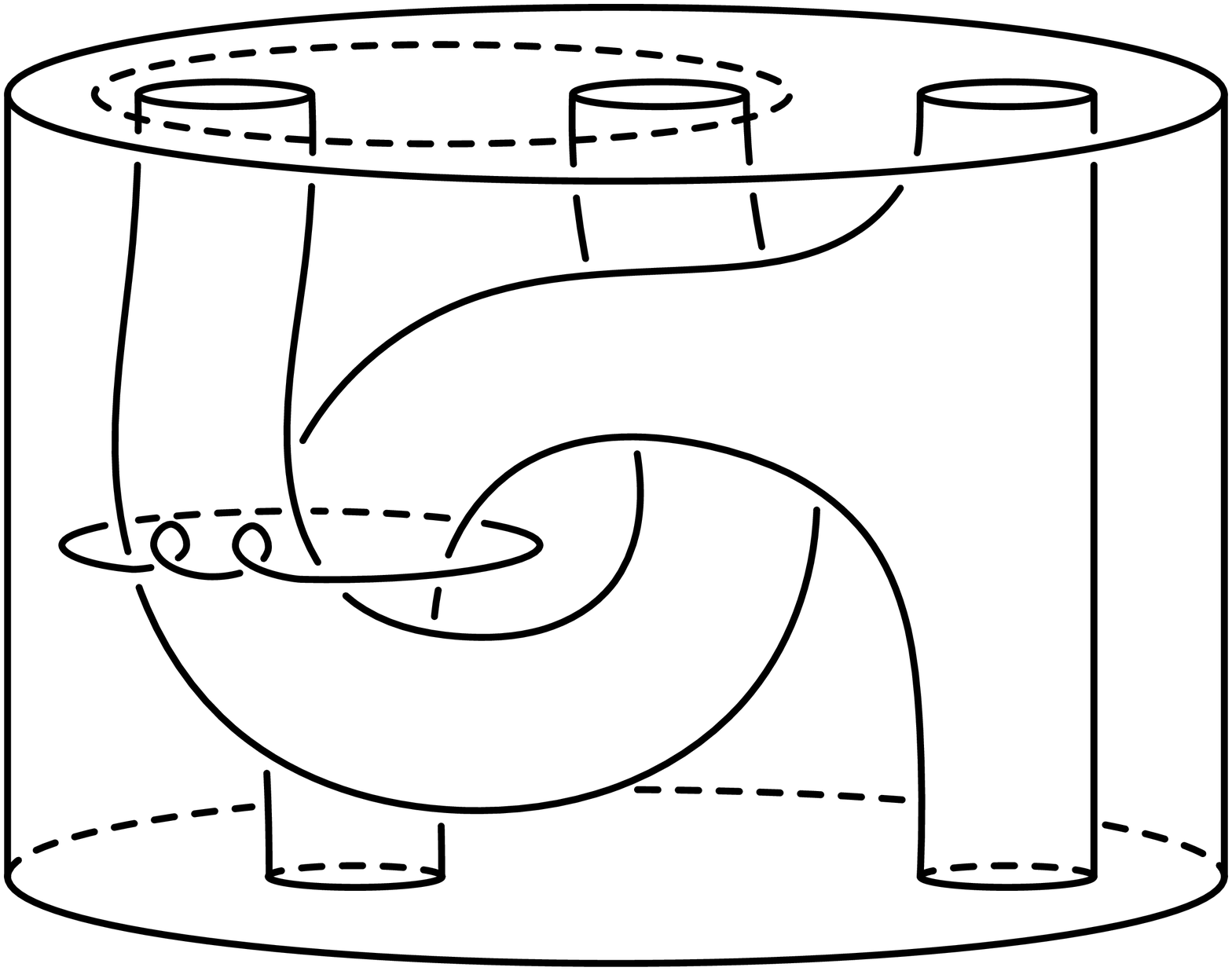}\vspace{0.75cm}
  \caption{Summary of $\pi/8$-phase gate protocol.}
  \label{fig:protocol_summary}
  \end{center}
\end{figure}

We now exhibit a topological protocol for using twisted interferometry to directly generate a $\pi/8$-phase gate, rather than by generating magic states (which are subsequently used to produce a $\pi/8$-phase gate). In comparison, this protocol has the advantage of being more efficient and not utilizing entangling gates. However, it requires that the twisted interferometry operation have sufficiently small errors, whereas the magic state generation protocol allows one to apply a high error threshold error-correction protocol, known as magic state distillation~\cite{Bravyi05}, if the twisted interferometry operation is not sufficiently free of error. The protocol described here, summarized in Fig.~\ref{fig:protocol_summary}, exhibits the roots of twisted interferometry in surfaces of positive genus. This protocol can be viewed as another translation of the $\pi/8$-phase gate protocol of Ref.~\cite{Bravyi00-unpublished}, which was developed in the series of papers~\cite{Freedman06a,FNW05b,Bonderson10}, in this case utilizing twisted interferometry.

%%%%%%%%%%%%%%%%%%%%%%%%%%%%%%%%%%%%%%%%%%%%%%%%%%%%%%%%%%%%%%%%%%%%%%%%%%%%%%%%%%%%%%%%%%%%%%%%%%%%%%%%%
%%%%%%%%%%%%%%%%%%%%%%%%%%%%%%%%%%%%%%%%%%%%%%%%%%%%%%%%%%%%%%%%%%%%%%%%%%%%%%%%%%%%%%%%%%%%%%%%%%%%%%%%%%

In Fig.~\ref{fig:protocol_summary}, the $t=0$ slice depicts a topological qubit partially encoded in two anti-dots, i.e. $S^1$ boundaries between the (spatial) system and vacuum. Each of the anti-dots/boundaries carries topological charge $\sigma$ and the $I$ and $\psi$ fusion channels of this pair comprise the qubit basis states. The first event (as time increases) is the creation of a new anti-dot (the local minima), which carries trivial topological charge $I$. At the saddle point, this anti-dot splits into two anti-dots (two $S^1$ boundaries between the system and vacuum), each of which carries topological charge $\sigma$. This charge distribution is not random, so it must be controlled using appropriately tuned potential wells and/or local measurements of the topological charge on the anti-dots. The third object occurring in Fig.~\ref{fig:protocol_summary}, is the twisted interferometric loop $\gamma$. By Section~\ref{sec:double_twist}, $\gamma$ will carry an $\omega_a$, depending on the twisted interferometry measurement outcome $a=I$ or $\psi$. In other words, this indicates which of the two types of Dehn surgery has been done on $\gamma$. The fourth event is a fusion of the pair of $\sigma$ charged anti-dots of the original qubit into a single anti-dot with topological charge $\alpha = I$ or $\psi$, which are equal probability outcomes of the fusion. The fifth event is a topological charge measurement of the charge $\alpha$, which can be measured by ordinary quasiparticle interferometry or a local energetic measurement. In the case when the measurement outcome is $\alpha=\psi$, an addition final step, not shown in Fig.~\ref{fig:protocol_summary} to avoid excessive clutter, is needed, wherein the anti-dot/boundary carrying charge $\alpha=\psi$ is fused/merged with one of the final anti-dots/boundaries carrying charge $\sigma$. This is necessary for the final system topological charge configuration to match the initial configuration. In other words, the final qubit state is (partially) encoded by the two charge $\sigma$ boundaries (contained within the dashed circle) on the $t=1$ surface, but, if $\alpha =\psi$, then this final step is necessary for it to be encoded in the same manner as it was at $t=0$.

The initial state $\ket{\Psi}$ at $t=0$ transforms into the final state $\ket{\Psi^{\prime}}= U(a,\alpha) \ket{\Psi}$ at time $t=1$, where the operator $U(a,\alpha)$ depends on the twisted interferometry measurement outcome $a=I$ or $\psi$ (i.e. the label $\omega_a$ on curve $\gamma$) and the measurement outcome of the topological charge $\alpha$. Using standard techniques of quantum topology, we will verify that the (single-qubit) operator $U(a,\alpha)$ acting on this topological qubit is given (up to insignificant overall phases) by
\begin{eqnarray}
\label{eq:U_II}
U(I,I) &=& U(\psi,I) =
\left[
\begin{array}{cc}
1 & 0 \\
0 & e^{-i \pi/4}
\end{array}
\right] = R_{-\frac{\pi}{4}} = R_{-\frac{\pi}{2}} R_{\frac{\pi}{4}} \\
\label{eq:U_psipsi}
U(I, \psi) &=& U(\psi, \psi) =
\left[
\begin{array}{cc}
1 & 0 \\
0 & e^{- i 3\pi/4}
\end{array}
\right] = R_{-\frac{3\pi}{4}}= R_{-\pi} R_{\frac{\pi}{4}}
\end{eqnarray}
Clearly, these are all related to the $\pi/8$-phase gate $R_{\frac{\pi}{4}}$ by a single-qubit Clifford gate, which may be generated using braiding transformations of Ising $\sigma$ quasiparticles.

%%%%%%%%%%%%%%%%%%%%%%%%%%%%%%%%%%%%%%%%%%%%%%%%%%%%%%%%%%%%%%%%%%%%%%%%%%%%%%%%%%%%%%
\begin{figure}[t!]
  \labellist \small\hair 2pt
  \pinlabel $\alpha$ at 578 380
  \pinlabel $\beta$ at 360 60
  \endlabellist
  \centering
  \includegraphics[width=3in]{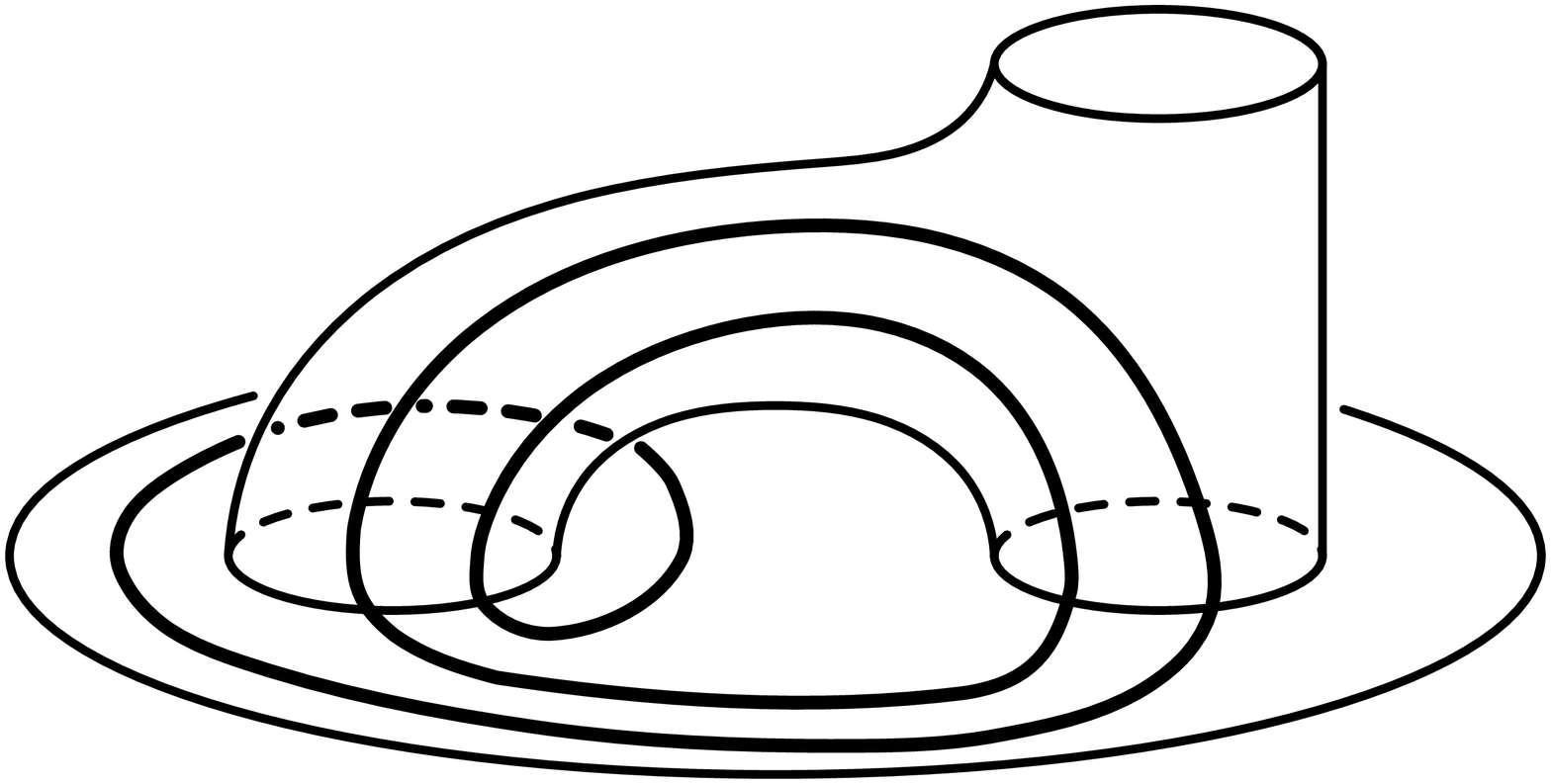}
  \caption{Cutting the (boundary) surface along $\beta$.}
  \label{fig:beta_cutting}
\end{figure}
%%%%%%%%%%%%%%%%%%%%%%%%%%%%%%%%%%%%%%%%%%%%%%%%%%%%%%%%%%%%%%%%%%%%%%%%%%%%%%%%%%%%%%

\begin{figure}[t!]
  \labellist \small\hair 2pt
  \pinlabel $\beta^{\prime}$ at 130 225
  \pinlabel $\mu$ at 330 580
  \pinlabel $\alpha$ at 545 575
  \pinlabel $\gamma^{\prime}$ at 285 180
  \endlabellist
  \centering
  \includegraphics[width=3in]{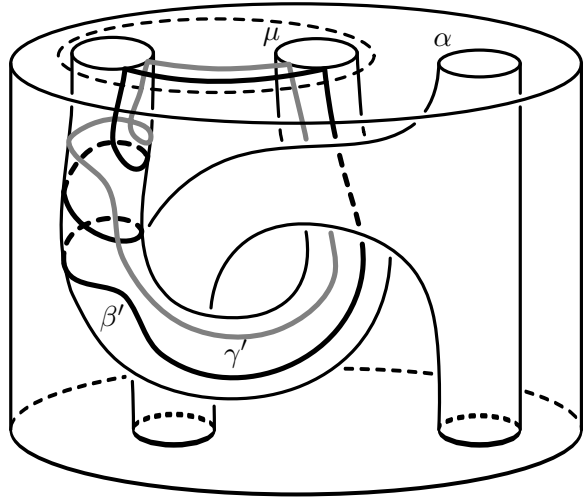}
  \caption{Thickening the surface from Fig.~\ref{fig:beta_cutting}.}
  \label{fig:thicken_surface}
\end{figure}

As seen in Refs.~\cite{Freedman06a,FNW05b,Bonderson10}, the $-\pi/8$-phase gate $R_{-\frac{\pi}{4}}$ is obtained, between the geometrically distinct initial and final ``marked pants,'' by cutting the surface open along $\beta$ in Fig.~\ref{fig:beta_cutting} if topological charge $\alpha = I$, and its inverse $R_{\frac{\pi}{4}}$ (the $\pi/8$-phase gate) if $\alpha = \psi$. Thickening the surface in Fig.~\ref{fig:beta_cutting} results in Fig.~\ref{fig:thicken_surface}. Now the framed curve $\gamma$ in Fig.~\ref{fig:protocol_summary} is precisely the surgery required to send $\beta$ to the meridian $\mu$ labeled in Fig.~\ref{fig:thicken_surface}. In both cases, the twisted interferometry measurement outcome $a=I$ effects ordinary framed surgery, while measuring $a=\psi$ effects a variant in which the core of the replacement solid torus carries a $\psi$-charge. The matrices in Eqs.~(\ref{eq:U_II})-(\ref{eq:U_psipsi}) give the precise gates $U(a,\alpha)$ executed according to the two outcomes $a$ and $\alpha$. Since the original qubit has $\sigma$ charges on its internal punctures, there will also be a $\sigma$-charge on $\beta$ (see Fig.~\ref{fig:beta_cutting}), but compared to the original qubit at time $t=0$, the relative phase between the two fusion channels $I$ and $\psi$ is now changed.

%%%%%%%%%%%%%%%%%%%%%%%%%%%%%%%%%%%%%%%%%%%%%%%%%%%%%%%%%%%%%%%%%%%%%%%%%%%%%%%%
%%%%%%%%%%%%%%%%%%%%%%%%%%%%%%%%%%%%%%%%%%%%%%%%%%%%%%%%%%%%%%%%%%%%%%%%%%%%%%%%%

The loop $\beta^{\prime}$ in Fig.~\ref{fig:thicken_surface} is simply a copy of $\beta$ transported across the product structure, i.e. through the topologically trivial $(2+1)$D spacetime bulk from one boundary surface to another. A $-1$ Dehn twist on the loop $\gamma^{\prime}$ throws $\beta^{\prime}$ to the meridian $\mu$. Thus, Dehn surgery on a torus in the bulk parallel to $\gamma^{\prime}$, with a $-1$ additional twist in its framing compared to the normal framing of $\gamma^{\prime}$ inherited from the boundary of the bulk, endows the bulk with a new product structure in which $\beta$ is connected by a cylinder to the meridian $\mu$. The curve $\gamma$, as drawn in Fig.~\ref{fig:protocol_summary}, is this additionally $-1$ framed bulk loop isotopic to $\gamma^{\prime}$. Thus, twisted interferometry with outcome $a=I$, in a sense, ``teleports'' the state from the non-time-slice qubit defined by cutting the surface of Fig.~\ref{fig:beta_cutting} along $\beta$ to the ``untwisted'' time-slice qubit defined by the top surface of Fig.~\ref{fig:thicken_surface}.

It remains to compute the effect of this protocol if the twisted interferometry measurement outcome is $a=\psi$. (Note: $a=\sigma$ is not a possible outcome as the charge along $\gamma = l^{\prime} = (1,-2)$ is obtained from the charge along $l$, which is initially in the $\{ I,\psi \}$ sector, by applying the matrix $B=S T^2 S^{-1}$, which does not mix the $\{ I,\psi \}$ sector and the $\sigma$ sector of the charge along $l$.) The effect of outcome $a=\psi$ will be a Wilson loop $\gamma^{\prime\prime}$ of charge $\psi$ parallel to $\gamma^{\prime}$ (in the bulk) with no additional twist in its framing.

Using the diagrammatic rules of UMTCs, we see that the effect of the protocol on the topological qubit basis states $q=I$ and $\psi$ is given by
\begin{eqnarray}
\pspicture[shift=-2](0.5,-2.2)(3,2.6)
\small
  \psset{linewidth=0.9pt,unit=0.5,linecolor=black,arrowscale=1.5,arrowinset=0.15}
  \psline(2,-1.5)(2,0.5)
  \psellipse(4,0.5)(1.5,2.25)
  \psellipticarc(-0.5,0)(1.03,0.53){-90}{0}
  \psellipticarc[border=1.5pt](0.5,0)(1.03,0.53){180}{270}
  \psarc(0,0){0.5}{0}{180}
  \psline(0.5,-0.5)(0.75,-0.5)
  \psellipticarc(0.75,0)(1.03,0.53){-90}{0}
  \psellipticarc[border=1.5pt](1.75,0)(1.03,0.53){180}{270}
  \psarc(1.25,0){0.5}{0}{180}
  \psarc(-0.5,0.5){1}{90}{270}
  \psarc[border=1.5pt](1.75,0.5){1}{-90}{90}
  \psline(-0.5,1.5)(1.75,1.5)
  \pscurve[border=1.5pt](2.5,4.25)(2.5,2.75)(4,1.25)(4,0.5)(2.5,-1)(2.5,-1.5)
  \pscurve(4,2.75)(4,3.25)(2.5,3.75)
  \psarc(2.25,-1.5){0.25}{180}{360}
  \psline[border=1.5pt](2,0.5)(2,4.25)
  \psellipticarc[border=1.5pt](4,0.5)(1.5,2.25){100}{180}
  \psline(4,-1.75)(4,-2.5)
  \psline{->}(-0.499,-0.4998)(-0.498,-0.5)
  \rput[bl](-1.125,-1.125){$\omega_a$}
  \rput[bc](2,4.5625){$\sigma$}
  \rput[bc](2.5,4.5625){$\sigma$}
  \rput[bl](4.03,3.28){$\alpha$}
  \rput[cr](1.7,-1.5){$\sigma$}
  \rput[cl](2.8,-1.5){$\sigma$}
  \rput[tc](4,-2.9){$q$}
  \rput[cc](3.35,-1.925){$\sigma$}
  \rput[cc](4.65,-1.925){$\sigma$}
\endpspicture
&=&
\sum_{z=I,\psi} C_{a,z}
\pspicture[shift=-2](0.5,-2.2)(3,2.6)
\small
  \psset{linewidth=0.9pt,unit=0.5,linecolor=black,arrowscale=1.5,arrowinset=0.15}
  \psline(2,-1.5)(2,-0.3)
  \psellipticarc(4,0.5)(1.5,2.25){-150}{130}
  \pscurve[border=1.5pt](2.5,4.25)(2.5,2.75)(4,1.25)(4,0.5)(2.5,-1)(2.5,-1.5)
  \pscurve(4,2.75)(4,3.25)(2.5,3.75)
  \psarc(2.25,-1.5){0.25}{180}{360}
  \psline[border=1.5pt](2,1.875)(2,4.25)
  \psellipticarc[border=1.5pt](4,0.5)(1.5,2.25){100}{130}
  \psline(4,-1.75)(4,-2.5)
  \rput[bc](2,4.5625){$\sigma$}
  \rput[bc](2.5,4.5625){$\sigma$}
  \rput[bl](4.03,3.28){$\alpha$}
  \rput[cr](1.7,-1.5){$\sigma$}
  \rput[cl](2.8,-1.5){$\sigma$}
  \rput[tc](4,-2.9){$q$}
  \rput[cc](3.35,-1.925){$\sigma$}
  \rput[cc](4.65,-1.925){$\sigma$}
  \pscurve(2.85,1.875)(2.375,1.125)(2,1.875)
  \pscurve(2.63,-0.3)(2.375,0.375)(2,-0.3)
  \psline(2.375,0.375)(2.375,1.125)
  \rput[cr](2,0.75){$z$}
\endpspicture
\notag \\
&=&
\sum_{z=I,\psi} C_{a,z} (-1)^{zq + z\alpha + z + \alpha q} e^{i \pi /8} R^{\alpha \sigma }_{\sigma} \left[R^{\sigma \sigma}_{q}\right]^{-1}
\pspicture[shift=-0.7](-0.8,-0.3)(1.3,1.3)
\small
  \psset{linewidth=0.9pt,linecolor=black,arrowscale=1.5,arrowinset=0.15}
  \psline(0,0)(0,0.5)
  \rput[cl](0.175,0.25){$q$}
  \psarc(0,1){0.5}{-180}{0}
  \rput[bc](-0.5,1.175){$\sigma$}
  \rput[bc](0.5,1.175){$\sigma$}
\endpspicture
\notag \\
&=&
[U(a,\alpha)]_{q,q}
\pspicture[shift=-0.7](-0.8,-0.3)(1.3,1.3)
\small
  \psset{linewidth=0.9pt,linecolor=black,arrowscale=1.5,arrowinset=0.15}
  \psline(0,0)(0,0.5)
  \rput[cl](0.175,0.25){$q$}
  \psarc(0,1){0.5}{-180}{0}
  \rput[bc](-0.5,1.175){$\sigma$}
  \rput[bc](0.5,1.175){$\sigma$}
\endpspicture
\end{eqnarray}
where $C_{I,I} = C_{\psi,\psi} =\cos (\pi/8)$ and $C_{I,\psi} = C_{\psi,I} = i \sin (\pi/8)$ are the coefficients resulting from the twisted interferometry with $(-2,0)$ twisting and outcome $a$. When the topological charge values $I$ and $\psi$ are written in the exponent, they are taken to mean $0$ and $1$, respectively. The coefficients $[U(a,\alpha)]_{q,q}$ in the final line are the diagonal elements of the unitary matrices $U(a,\alpha)$ (up to unimportant overall phases, i.e. phases that are independent of $q$) given in Eqs.~(\ref{eq:U_II}) and (\ref{eq:U_psipsi}). It is clear from the diagrams that the off-diagonal elements of the operators generated by this protocol must vanish, by conservation of topological charge.

\section*{Acknowledgements}

We thank R. Lutchyn, C. Nayak, K. Shtengel, and J. Slingerland for illuminating discussions. P.~B. and M.~F. thank the Aspen Center for Physics for hospitality and
support under the NSF Grant No. 1066293.

%\bibliography{corr}
%\bibliographystyle{elsart-num}

\end{document}